\documentclass[aps,twocolumn,prc,amssymb,amsmath,superscriptaddress, bm, floatfix,10pt]{revtex4-1}
\usepackage{mathptmx}
\usepackage{graphicx}
\usepackage{color}
\makeindex

\begin{document}

\title{Surface energy of magnetized superconducting matter in the neutron star cores}
\author{D. N. Kobyakov}
\email{dmitry.kobyakov@appl.sci-nnov.ru}
\affiliation{Institute of Applied Physics of the Russian Academy of Sciences, 603950 Nizhny Novgorod, Russia}
\date{9 October 2020}

\begin{abstract}
In the outer core of neutron stars (NS) quantum liquids are thought to exist during most of the observable stellar lifetime.
These liquids are characterized by two different, density-dependent superfluid (SF) condensation critical temperatures associated with neutrons and protons.
A possibility that cannot be excluded in the state-of-the-art astrophysics, and that warrants examination, is a spatial overlap of the SF and superconducting (SC) domains.
Partial overlap would imply that there is a layer where both critical temperatures are equal.
In vicinity of this layer the Ginzburg-Landau (GL) effective field theory is valid.
In this paper, an effective field theory for proton superconductor (SC) interacting with neutron SF, both with scalar order parameters, is developed and applied to the surface energy (SE) of a magnetized SC body.
Essentially, the SE studied here differs from the nuclear SE: here, the proton SF density decays to zero while the total proton density is constant across the surface.
Interactions between the condensates are parameterized phenomenologically and their effects determined from calculations of a planar SE as the ranges of parameters are varied.
The critical GL parameter $\kappa_c$ which renders the SE equal to zero is found analytically by noting that in a system with vanishing SE the thermodynamic critical MF is equivalent to the upper critical MF.
In the case of weak coupling, $\kappa_c$ is shown to be a linear function of SF-SF density-density coupling, in agreement with the earlier results based on asymptotic intervortex interactions.
Numerical simulations corroborate our analytical predictions.
Coupling due to the mixed term arising from a scalar product of gradients of the SF densities, which had been considered in the earlier literature, is seen to have practically no effect on the superconductivity type.
However, this coupling does produce a frozen wave packet of the SF neutron density localized at the surface.
It is shown that the leading contribution from the gradient coupling arises from a novel mixed quantum pressure term, but still does not affect the planar SE.
The present calculations provide an initial map of superconductivity types in the phenomenological effective field theory and will serve as a landmark for future studies, which require microscopic calculations of the coupling parameters introduced here phenomenologically.
\end{abstract}
\maketitle

\section{Introduction}
Macroscopic quantum phenomena in superconducting neutron stars are directly linked to their observable signatures.
For example, protons superconductivity is expected in the core of NS and is a crucial problem in many aspects of rotational and magnetic evolution of pulsars and magnetars.
The response of matter to the magnetic field is a fundamental property and it is conveniently expressed in terms of superconductivity type.
The type of superconductivity relevant to uniform nuclear matter was first analyzed at typical conditions in the core of NS by Baym, Pethick and Pines (1969) in the framework of a single-component superconductor (influence of SF neutrons was discarded) \cite{BaymPethickPines1969}.
The conclusion was that type-II is expected because the GL parameter in NS matter is around 3, while the critical GL parameter $\kappa_c$ for a single-component SC is equal to $1/\sqrt{2}$.
However, the SF state in the NS interior is expected to have more than one scalar (proton) condensate and it is important to understand multicomponent SF effects arising from the nuclear interactions.
Coupling of the stellar angular momentum to the MF is set by the vortex-flux tube interactions (or their absence) \cite{AlparLangerSauls1984,RudermanZhuChen1998,Link2003}.
The magnetic stresses and arrangement of the MF lines in the NS interior are also influenced by the core superconductivity \cite{EassonPethick1977,Lander2013}.

Link (2003) \cite{Link2003} argued that pulsar observations suggest there is either no coexisting proton type-II superconductivity and neutron superfluidity, or there is type-I proton superconductivity which might coexist with SF neutrons.
Buckley, Metlitski and Zhitnitsky (2004) \cite{BMZ2004} suggested that a SF-SC density-density coupling between the protons and the neutrons might be capable of changing the type of the proton superconductivity.
Alford, Good and Reddy (2005) \cite{AlfordGoodReddy2005}, using a simple model of a dilute Fermi gas with two-body contact interactions, argued that such a coupling vanishes due to a large difference between the Fermi momenta of neutrons and protons.
However, the nucleon interactions are strong and this is a simple reason why models suitable for dilute systems with perturbative interactions in general fail to represent the basic physical features of nuclear matter.

The intrinsic three-body fermion interaction might naturally cause the SF-SC density-density coupling irrespective of the difference of the Fermi momenta, because it couples the isospin degrees of freedom in the differential equations for normal and anomalous propagators of the fermions.
This qualitative argument is one of motivations for the present work.
The standard machinery of the propagator equations in a single-component SC (with only one isospin component) is given in the classical work of Gorkov \cite{Gorkov1958}.
It is straightforward to generalize this approach to the two-component mixture considered here \cite{Kobyakov2019unp}, however further work is necessary because three-body interactions in nuclear matter are characterized by a finite range.

Another motivation is to provide an independent investigation of the problem of parametric transitions between types of superconductivity, which has been considered earlier from the point of view of the vortex-vortex asymptotic interactions \cite{AlfordGood2008}.
Here, a flat interface is employed and the associated surface energy is computed.
This approach, though is explicitly very different, is expected to provide equivalent results compared with the approach of intervortex interactions used by Alford \& Good (2008) \cite{AlfordGood2008}.

A standard framework for describing the physics of SCs is a phenomenological GL model \cite{GinzburgLandau1950}.
In NS matter there are two SF species characterized by two different transition temperatures, representing a logical obstacle for a direct extension of the GL model to the two-component mixture.
In the case where there are several unequal SF critical temperatures, then the validity of the "GL" expansion of the condensation energy simultaneously fails for the both components, because the condensate with a higher critical temperature is not small in magnitude at temperatures slightly below the lower critical temperature.
The GL model is applicable near the critical temperature, and the basic assumption is that the free energy can be expanded in powers of the order parameter, thus fixing the form of the SF energy functional \cite{LL1980}.
In order to avoid ambiguities one should abandon the terminology "GL model at zero temperature", because the expansion of the GL free energy in powers of the small order parameter is valid only near the critical temperature.
At zero temperature, the magnitude of the order parameter is not small.
These difficulties are overcome by considering the system near the phase transition temperature and by assuming that the critical temperatures are equal.
Since the gaps are smooth functions of the total baryon density in uniform nuclear matter, this assumption is realistic if the SF and the SC spatial domains overlap in the stellar interior.
Indeed, in this case there must be a surface within the overlapping region where the critical temperatures are equal.

Here, Cooper pairing in the $S$-wave channel is assumed for both protons and neutrons, although this is a simplified picture because the $S$-wave neutron gap is expected to close for relevant nuclear matter densities.
A more realistic picture would include higher partial waves for neutron quasiparticles and self-consistently account for unequal critical temperatures for neutrons and protons.
In fact, even symmetry of the neutron SF order parameter is presently an open question.
It has been shown that the tensor component of the two-body quasiparticle interaction and the $P$-wave SF neutron gap are significantly reduced due to the medium polarization corrections, in particular due to the coupling of the tensor and the spin-orbit force to the strong spin-spin interaction \cite{SchwenkFriman2004}.
However, the three-body forces and higher many-body corrections are expected to provide significant effects and need to be included in future quantitative studies of the gaps \cite{ZhouEtal2004,DrischlerEtal2017}.

Presently, the presence and exact distribution of superconducting domains in the interior of NS is an open problem.
Even a small layer exhibiting type-I superconductivity, or its absence, could have relevant astrophysical consequence because it would affect the crust-core boundary conditions for the MF and hence will provide significant constraints for theoretical models of the stellar MF and the related phenomena.
However, low-energy constants of the model are linked to the low-energy constants of the effective theories of nuclear interactions in a non-trivial way.
The general wisdom suggests that NS matter can be only type-II.
Nevertheless, the development of an effective field theory and systematical study of various effective interactions is required and will be the focus here.

The present two-component model provides insight into superconducting properties of systems with several components and should be investigated prior to development of more sophisticated models.
The Fermi momenta of protons and neutrons are very different as a result of nuclear equilibrium \cite{BaymBethePethick1971}.
The order parameter is diagonal in the isospin quantum number because particles with significantly different Fermi momenta cannot form a Cooper pair with zero total momentum.
Some aspects of the two-component GL model in the context of the NS matter were studied earlier \cite{AlparLangerSauls1984,RudermanZhuChen1998,BMZ2004,BuckleyEtalPRC2004,Babaev2004,AlfordGood2008,DrummondMelatos2017,HaberSchmitt2017}.
The vector coupling of the SF components was represented in the free energy functional by "\emph{current} $\cdot$ \emph{current}" terms \cite{AlparLangerSauls1984,DrummondMelatos2017} or by "\emph{SF density gradient} $\cdot$ \emph{SF density gradient}" contributions \cite{AlfordGood2008,HaberSchmitt2017}.
Here, both types of coupling are retained and are called simply "momentum-coupling" and "gradient-coupling", correspondingly.
It is convenient to introduce a shorthand term "density-coupling" that refers to the SF-SC density-density coupling between the protons and neutrons.
In this paper, a novel SF interaction is considered in the form of a "\emph{gradient of wave function modulus} $\cdot$ \emph{gradient of wave function modulus}", which represents a mixed quantum pressure contribution.

The plan of the paper is as follows.
Section II systematically introduces the model.
Section III provides the basic equations of the model and the parameterization.
Also, uniform mixtures are considered and thermodynamic stability conditions are displayed in Fig. 1 for a special case of equal quartic GL parameters.
Finally, the setup for finding planar interfacial profiles and the surface energy in the magnetic field is established.
Section IV discusses types of superconductivity and gives derivations of the main analytical results using a novel method to find the critical GL-parameter.
Section V provides the mathematical definition of the surface energy of a planar interface, which is used in the numerical calculations.
Section VI presents a description of the numerical procedure.
The main numerical results are shown in Table I and in Figs. 2-3.
Conclusions are given in Sec. VII.
The appendix presents some useful results from the Sch$\ddot{\mathrm{o}}$dinger model.

\section{Phenomenology}
\subsection{Effective field in SF-SC system}
Any physical theory is phenomenological, the only difference is the energy scale involved in the problem.
Phenomenology of the present model is understood as the low-energy effective-field description of SF and SC matter, on length scales that resolve the spatial variations of the order parameter.
It involves quasiclassical fields of the SF-SC order parameter and the MF in the Coulomb gauge.
The field interactions are given by the effective couplings and do not explicitly involve the quantum interaction fields.
The present phenomenology is an effective low-energy theory for quantum degenerate uniform nuclear matter.
There are two important limiting cases for the SF phenomenology: either just below the temperature-driven SF phase transition point -- the GL model, or at ultra low temperature -- ideal fluid hydrodynamics.

The effective degrees of freedom in the present theory represent the SF and SC properties of the system.
The SF-SC phase physically differs from  normal matter because its symmetry is different.
This situation has some analogies with the usual water-ice phase transition, which is a representative of the first-order phase transition.
The new phase appears within localized regions of space ("seeds") and is characterized by a nonzero enthalpy that is required to assemble or melt a seed with minimum size determined by the interplay of the surface and volume energies of the seed.
On the contrary, in second order phase transitions (which occur, for example, in the MF driven superconductive phase transition in type-II SCs) the new phase appears continuously in the whole body of the system and the enthalpy at the transition is zero.
This fundamental difference can be understood as a consequence of either positive or negative excess energy associated with surface of the new phase immersed in the normal phase.

This paper is focused on the structure of the order parameter variations between zero and the bulk values as a result of the MF screening by the Meissner currents, and therefore the phenomenology used here takes the view on mesoscopic rather than macroscopic (hydrodynamic) scales, while information on the hydrodynamic excitations is lost.
Indeed, the GL model for terrestrial SCs works well only for mesoscopic length scales and does not describe the hydrodynamic excitations.
Both the GL theory and the zero-temperature hydrodynamics are based on the notion of the wave function, but can be equivalently formulated in terms of the SF densities and velocities.
This duality is a convenient feature for application to various types of problems, for the wave function description is in fact a mechanical description in terms of densities and momenta as opposed to description in terms of densities and velocities.
Recently, it has been explicitly shown in detail \cite{Kobyakov2018},\cite{Gavassino2019} that treatment for the SF hydrodynamics in terms of fluid momenta \cite{KobyakovPethick2017} is equivalent to treatment in terms of velocities \cite{ComerJoynt2003},\cite{Prix2004}.

While the zero-temperature hydrodynamics of the SF-SC mixture is representable as a multicomponent Schr$\ddot{\mathrm{o}}$dinger equation, the GL theory is formulated only in terms of stationary wave function that satisfies a stationary Schr$\ddot{\mathrm{o}}$dinger equation.
In both cases, the wave function can be written in the Euler form
\begin{equation}\label{psi}
\psi_\alpha(t,\mathbf{r})\equiv\sqrt{n_\alpha(t,\mathbf{r})}e^{\mathrm{i}\phi_\alpha(t,\mathbf{r})},
\end{equation}
where $\alpha=p$ or $\alpha=n$ is the isospin index, $n_\alpha$ is the SF density, and $\phi_\alpha$ is the SF phase.
Physics of the model is contained in the action functional
\begin{equation}\label{Spsi}
S=\int dt \int d^3\mathbf{r}\quad n_\alpha\frac{\partial{\phi_\alpha}}{\partial t}-H[n_\alpha,\phi_\alpha],
\end{equation}
where $H[n_\alpha,\phi]$ is the total energy density.
\subsection{The total, SF and normal densities}
\label{TSNd}
Generally, the SF density is different from the total particle density (a well-known exception is given by ultracold dilute gases made up of integer-spin repulsive particles, where the SF density is to a good approximation equal to the total density).
The functions in the uniform bulk are marked by the additional subscript "0".
The total nucleon density $n^{\mathrm{tot}}_{\alpha 0}$ is a sum of the SF and the normal densities:
\begin{equation}\label{ntot}
n^{\mathrm{tot}}_{\alpha0}=n_{\alpha0}+n^{\mathrm{n}}_{\alpha0}.
\end{equation}

The density decomposition used in Eq. (\ref{ntot}) implies that there are normal and SF types of motion for both isospin species.
Note that the normal component is not necessarily a fluid (for example, in the inner crust of NS the normal component is solid-like and is significant even at zero temperature).
Each type of motion is associated with its own effective mass, which is temperature-dependent, and with a characteristic structure of the velocity field.
The SF density is associated with the corresponding macroscopic wave function $\psi_{\alpha0}$ in the uniform bulk \cite{Feynman1998}:
\begin{equation}\label{nSF0}
  n_{\alpha0}\equiv|\psi_{\alpha0}|^2.
\end{equation}
It is convenient to define an analogue of Eq. (\ref{nSF0}) generalized to non-uniform SF matter:
\begin{equation}\label{nSF}
  n_{\alpha}(t,\mathbf{r})\equiv|\psi_{\alpha}(t,\mathbf{r})|^2.
\end{equation}
The normal density is also a function of spatial variables:
\begin{equation}\label{nSF}
  n_{\alpha}^{\mathrm{n}}=n^{\mathrm{n}}_{\alpha}(t,\mathbf{r}).
\end{equation}
The SF component is described by two new independent variables -- the SF density $n_{\alpha}(t,\mathbf{r})$ and the SF phase $\phi_{\alpha}(t,\mathbf{r})\equiv (1/2\mathrm{i})\ln(\psi_{\alpha}/\psi_{\alpha}^*)$.
In a mixture, an important parameter is the ratio of the SF densities in the bulk which is denoted $\eta$:
\begin{equation}\label{etaDef}
  \eta\equiv\frac{n_{p0}}{n_{n0}}.
\end{equation}
At temperatures much lower than $T_c$, which is of the order of the SF energy gap $\Delta_\alpha$, the normal matter density is negligible and $n_{\alpha0}^{\mathrm{tot}}=n_{\alpha0}$ to a good approximation.
In this case, the basic assumption of the GL theory is violated, however, the parameter $\eta$ is directly related to the proton fraction $x_p$:
\begin{equation}\label{etaX}
  \eta=\frac{x_p}{1-x_p} \qquad \mathrm{at}\quad \frac{\Delta_\alpha}{k_BT}\gg1,
\end{equation}
where
\begin{equation}\label{xfrac}
  x_p\equiv\frac{n_{p0}^{\mathrm{tot}}}{n_{p0}^{\mathrm{tot}}+n_{n0}^{\mathrm{tot}}}.
\end{equation}
$k_B$ is the Boltzmann constant, and $\Delta_\alpha$ is the SF energy gap of baryons with isospin $\alpha$.

In the opposite case, when $T\sim T_c$, there is no direct relation between the parameters $\eta$ and $x_p$ because the former is defined by the the pairing interactions between the baryons while the latter is defined by the chemical reactions.
Nuclear matter is found to be strongly isospin asymmetric with $x_p\sim5\%$ \cite{BaymBethePethick1971}.
The SF density ratio $\eta$ is expected to take on values between 0 and $+\infty$.
\subsection{The power counting scheme}
In this paper, the interactions are thermodynamic and involve only classical fields (the vector potential).
One may assume that the magnitude of both energy and momentum excitations in the system are sufficiently small so that the effective field description on mesoscopic length scales is valid.
In the present effective theory the basic degrees of freedom of the system are the SF densities $n_\alpha$ and the phases $\phi_\alpha$, which represent the effective field for the system.
The order of magnitude of a contributing term of type $i$ is characterized  by a SF interaction index $\Delta_i$:
\begin{equation}\label{interactionIndex}
  \Delta_i=d_i+{n_i}-2,
\end{equation}
where $d_i$ is number of gradients and $n_i$ is sum of powers of the both SF densities.
When the temperature is close to the SF critical temperature, there is an additional expansion of the nonlinear contributions to the total free energy in terms of even powers of the order parameter modulus.

To leading order, the SF energy is given by the condensation energy.
In the GL theory it is expanded in powers of the SF density, and for a scalar order parameter this expansion is given by the quadratic form $n_\alpha n_\beta$.
Although a term of the form $\propto n_p n_n$ is a leading order one, the present effective theory does not specify the weighting factor for this term, which might turn out to be vanishingly small.

In the next-to-leading order, $\Delta_i=1$, contribution to the SF energy is provided by quasiclassical kinetic energies
\begin{equation}\label{NLOkin1}
\frac{1}{2 m}n_\alpha\mathbf{P}_\alpha^2
\end{equation}
and by the quantum pressure terms
\begin{equation}\label{NLOkin2}
\frac{\lambda_{\alpha\beta}}{2m}(\hbar\nabla\sqrt{n_\alpha})\cdot(\hbar\nabla\sqrt{n_\beta}).
\end{equation}
Coefficients of the diagonal terms are equal to $1$ in the GL theory:
\begin{equation}\label{lambdaDiag}
\lambda_{pp}=\lambda_{nn}=1,
\end{equation}
thus, the sum of the kinetic and the quantum pressure energies can be combined into a single quantum kinetic energy term: $\frac{1}{2}mn_\alpha\mathbf{P}_\alpha^2+\frac{1}{2}(\hbar\nabla\sqrt{n_\alpha})^2=\hbar^2|\nabla\psi_\alpha|^2/2m$.
The non-diagonal terms $\propto\lambda_{np}$ provide the leading correction for the interaction of the type "\emph{gradient of wave function modulus} $\cdot$ \emph{gradient of wave function modulus}":
\begin{equation}\label{NLOkin3}
\lambda_{np}(\hbar\nabla\sqrt{n_p})(\hbar\nabla\sqrt{n_n})/2m.
\end{equation}

In the next-to-next-to-leading order, the neutron-proton interaction corrections are characterized by $\Delta_i=2$ and involve the following terms:
\begin{eqnarray}
  && \label{NNLO1} n_\alpha n_\beta(\mathbf{P}_\alpha-\mathbf{P}_\beta)^2,\\
   && \label{NNLO2} \quad\quad\mu_{\alpha\beta}(\nabla n_\alpha)\cdot(\nabla n_\beta).
\end{eqnarray}
Terms of the form $\propto(\mathbf{P}_\alpha-\mathbf{P}_\beta)\cdot(\nabla{n_\alpha})$ are discarded on symmetry grounds because they violate the time-reversal symmetry since $\mathbf{P}_\alpha$ changes sign when the time is reversed.
Usually, the SF entrainment is parameterized by a bilinear functional of the SF densities $n_{np}=n_{np}[n_p,n_n]$ \cite{BorumandJoyntKluzniak1996,KobyakovEtal2017,Kobyakov2018}.
In this case, the entrainment is represented by contributions of the type indicated in Eq. (\ref{NNLO1}).

The term from Eqs. (\ref{NNLO2}) has been considered in the earlier literature \cite{AlfordGood2008}.
However, the magnitude of this term is smaller than the magnitude of the next-to-leading order correction given in Eq. (\ref{NLOkin3}), which has not been included in \cite{AlfordGood2008}.
Alford \& Good in \cite{AlfordGood2008} assumed
\begin{equation}\label{muAA}
\mu_{np}=\mathrm{const},\quad\mu_{pp}=\mu_{nn}=0.
\end{equation}
Expansion in powers of the excitation momenta $\mathbf{P}_\alpha$ is equivalent to expansion in powers of SF velocities $\mathbf{v}_\alpha$, because those are linearly linked.

The present effective theory is non-relativistic but it is instructive to succinctly describe origins of the relativistic corrections.
Evaluation of the baryon Fermi energies at typical conditions in the outer core of NS shows that the nucleons are non-relativistic to a good approximation.
The energy of the relativistic electrons is irrelevant for calculation of the SE.
Relativistic corrections to the kinetic energy of the matter flows contribute with a factor of $(\mathbf{v}_\alpha/c)^2$ and may be safely discarded.
Moreover, there is a dependence of the SF densities on the momenta due to the Fermi liquid effects, however those terms represent contributions with the SF interaction index $\Delta_i>2$ and may be neglected in the present model with $\Delta_i\leq2$.
\subsection{Energy of the effective field}
The excitation momentum associated with a low energy perturbation is linked to the phase field by the gradient operator: in neutral SFs $\mathbf{P}_\alpha=\hbar\nabla\phi_\alpha$, and in a mixture of neutrons and the protons
\begin{equation}\label{momentaDef}
\mathbf{P}_n=\hbar\nabla\phi_n,\quad \mathbf{P}_p=\hbar\nabla\phi_p-\frac{e}{c}\mathbf{A}.
\end{equation}
The phase of the proton wave function enters in the gauge-invariant form
\begin{equation}\label{gaugecoupling}
(\nabla\phi-\frac{e}{\hbar c}\mathbf{A}),
\end{equation}
where $e$ is the proton charge and $\mathbf{A}$ is the electromagnetic vector potential (a three-dimensional vector field).
Starting from an electrically neutral theory, the gauge coupling is provided by a substitution
\begin{equation}\label{Gcoupl}
  (\nabla\psi)\rightarrow(\hat{\mathbf{D}}\psi),
\end{equation}
where
\begin{equation}\label{Dhat}
\hat{\mathbf{D}}\equiv\nabla-\mathrm{i}\frac{e}{\hbar c}\mathbf{A},
\end{equation}
while $\nabla|\psi|$ remains $\nabla|\psi|$.

According to the power counting scheme, the structure of the perturbation energy of the two fluids with particles of the same mass $m$ written up to the next-to-next-to-leading order is the following:
\begin{eqnarray}
  &&\nonumber E^{\mathrm{tot}}=\int d^3\mathbf{r}\quad E_{\mathrm{st}}^{\mathrm{nuc}}[n^{\mathrm{n}}_{p},n^{\mathrm{n}}_{n},n_p,n_n]+\frac{(\nabla\times\mathbf{A})^2}{8\pi}\\
  &&\nonumber + n_p \frac{P_p^2}{2m} + n_n \frac{P_n^2}{2m} + \lambda_{pp}\frac{(\hbar\nabla\sqrt{n_p})^2}{2m} + \lambda_{nn}\frac{(\hbar\nabla\sqrt{n_n})^2}{2m} \\
  &&\nonumber + \lambda_{np}\frac{\hbar^2}{2m}(\nabla\sqrt{n_p})(\nabla\sqrt{n_n}) \\
  &&\nonumber - n_{np}\frac{(\mathbf{P}_p-\mathbf{P}_n)^2}{2m}+\frac{\mu_{np}}{4}\left(\nabla n_p\right)\cdot\left(\nabla n_n\right)\\
  &&\label{Emom}+\frac{\mu_{pp}}{4}\left(\nabla n_p\right)^2+\frac{\mu_{nn}}{4}\left(\nabla n_n\right)^2.
\end{eqnarray}
The fluid number currents
\begin{equation}\label{currdef}
\mathbf{J}_\alpha=n_\alpha \mathbf{v}_\alpha
\end{equation}
are obtained from the definition
\begin{equation}\label{curr2def}
  \mathbf{J}_\alpha=\frac{\delta E^{\mathrm{tot}}}{\delta \mathbf{P}_\alpha},
\end{equation}
and no summation over repeated indices is implied in this paper.
The velocities $\mathbf{v}_\alpha$ are related to the momenta $\mathbf{P}_\alpha$ by a linear transformation that preserves the Galilean invariance
\begin{eqnarray}
\label{vel1} && \mathbf{v}_p=\frac{n_{pp}\mathbf{P}_p+n_{np}\mathbf{P}_n}{mn_p},\\
\label{vel2} && \mathbf{v}_n=\frac{n_{nn}\mathbf{P}_n+n_{np}\mathbf{P}_p}{mn_n},
\end{eqnarray}
where $n_{\alpha\alpha}=n_{\alpha}-n_{np}$.
The velocity difference is proportional to the momenta difference: $(\mathbf{v}_p-\mathbf{v}_n)\propto(\mathbf{P}_p-\mathbf{P}_n)$.
For each of the fluids, the individual kinetic energy density written in terms of the velocity has the standard form: $mn_\alpha v_\alpha^2/2$.

An equivalent expression for the perturbation energy in terms of velocities reads
\begin{eqnarray}
  \nonumber&& E^{\mathrm{tot}}=\int d^3\mathbf{r}\quad mn_p v_p^2/2 + mn_n v_n^2/2 \\
  \label{Evel}&& + U[n_\alpha,(\mathbf{v}_p-\mathbf{v}_n)^2]+\frac{(\nabla\times\mathbf{A})^2}{8\pi},
\end{eqnarray}
where the internal fluid energy density $U[n^{\mathrm{n}}_{\alpha},n_\alpha,(\mathbf{v}_p-\mathbf{v}_n)^2]$ is a functional of the normal baryon densities, the SF densities, and the square of the SF velocity difference $(\mathbf{v}_p-\mathbf{v}_n)^2$.
The Galilean invariance  is the reason why $U$ is a functional of $(\mathbf{v}_p-\mathbf{v}_n)^2$: the internal fluid energy is the same in any inertial frame of reference.
In the absence of matter flows and density gradients, the uniform nuclear energy $E_{\mathrm{st}}^{\mathrm{nuc}}$ is equal to the internal energy density $U$:
\begin{equation}
\label{EstU}
E_{\mathrm{st}}^{\mathrm{nuc}}[n^{\mathrm{n}}_{\alpha},n_\alpha]=U[n^{\mathrm{n}}_{\alpha},n_\alpha,(\mathbf{v}_p-\mathbf{v}_n)^2=0,\nabla n_\alpha=0].
\end{equation}

The Hamiltonian of the ideal fluid mixture is valid in the limit of zero temperature is characterized by $\lambda_{\alpha\beta}=\mu_{\alpha\beta}=0$.
In terms of the macroscopic wave functions $\psi_\alpha$ and $\psi_\alpha^*$, it reads:
\begin{eqnarray}
  \nonumber &&H^{\mathrm{id}}=\frac{\hbar^2}{2m}\left[|\hat{\mathbf{D}}\psi_p|^2-(\nabla|\psi_p|)^2\right]+\frac{\hbar^2}{2m}\left[|\nabla\psi_n|^2-(\nabla|\psi_n|)^2\right]\\
  \nonumber &&-\frac{\hbar^2}{2m}n_{np}\left[\frac{1}{2\mathrm{i}} \left( \frac{\psi_p^*\hat{\mathbf{D}}\psi_p-c.c.}{n_p} - \frac{\psi_n^*\nabla\psi_n-c.c.}{n_n}  \right) \right]^2\\
  \label{Emom}&& + E_{\mathrm{st}}^{\mathrm{nuc}}[n^{\mathrm{n}}_{p},n^{\mathrm{n}}_{n},n_p,n_n].
\end{eqnarray}
The quantum pressure contributions $(\nabla|\psi_\alpha|)^2$ are subtracted (since $\lambda_{\alpha\beta}=0$) and thus are removed from the energy functional because the ideal fluid hydrodynamics at zero temperature has no information on the short length scale phenomena such as the vortex core structure.
\subsection{SF-SF entrainment and the other next-to-next-to-leading order corrections}
The entrainment contribution to the total energy is the next-to-next-to-leading order correction and it arises as a cross term in momenta and has the form $n_{np}[n_p,n_n](\mathrm{\mathbf{P}}_p-\mathrm{\mathbf{P}}_n)^2$ which is Galilean-invariant.
For problems where the electromagnetic coupling is irrelevant it is convenient to rewrite the momentum-coupling in the following compact form:
\begin{eqnarray}
\nonumber&&
-\frac{\hbar^2}{2m}n_{np}\left[\frac{1}{2\mathrm{i}} \left( \frac{\psi_p^*\nabla\psi_p-c.c.}{n_p} - \frac{\psi_n^*\nabla\psi_n-c.c.}{n_n}  \right) \right]^2 \\
\label{Eentr}&&  = -\frac{\hbar^2}{2m}n_{np}\left(\frac{1}{2\mathrm{i}}\nabla \ln\frac{\psi_p\psi_n^*}{\psi_p^*\psi_n}\right)^2.
\end{eqnarray}
Dissipationless entrainment of one SF by a flow of another SF in the mixture can be viewed as the isospin mixing term with one pair of the gradient operators (if the gradient applies to the full complex-valued proton field, the gauge-invariant spatial derivative is implied):
\begin{eqnarray}\nonumber&&
 \!\!\!\!\!\!\!\!\!\!\!\!\!\!\!\!\!\!\!\!\!\!\!\!\sum_{\begin{array}{c}
           ^{j,k=x,y,z} \\
           ^{\alpha,\beta,\gamma,\delta=p,n} \\
         \end{array}
 } \left[ c^{jk}_{\alpha\beta\gamma\delta}{\psi}_\alpha{\psi}_\beta^*(\nabla_j{\psi}_\gamma)(\nabla_k{\psi}_\delta^*)\right.\\
\label{order4terms}&&\qquad\qquad\quad\quad\left.+d^{jk}_{\alpha\beta\gamma\delta}{\psi}_\alpha{\psi}_\beta(\nabla_j{\psi}_\gamma^*)(\nabla_k{\psi}_\delta^*)\right].
\end{eqnarray}
This expression may be simplified because the coefficients $c^{jk}_{\alpha\beta\gamma\delta}$ and $d^{jk}_{\alpha\beta\gamma\delta}$ are constrained by the Galilean invariance of the internal energy, and by the isotropy of space and by the time-reversal symmetry of the system.
In order to satisfy the symmetry requirements, the coefficients $c^{jk}_{\alpha\beta\gamma\delta}$ and $d^{jk}_{\alpha\beta\gamma\delta}$ must be functionals of the SF densities as we shall see below.
\subsection{A "Ginzburg-Landau" model for a mixture with equal critical temperatures}
Following the notation of  \cite{AlparLangerSauls1984} I introduce $f_u[n_p,n_n]$ -- the condensation energy density, which is a functional of the SF densities $n_\alpha$.
Adopting also the notation from \cite{LL1980} I introduce the condensation energy via the GL expansion
\begin{eqnarray}
\nonumber  && \!\!\!\!\!\!\!\!\!\!\!\!\!\!\!\!\!\!\!\!f_u=a_p|\psi_p|^2+a_n|\psi_n|^2\\
\label{fu}  && +\frac{b_{p}}{2}|\psi_p|^4+\frac{b_{n}}{2}|\psi_n|^4+b_{np}|\psi_p|^2|\psi_n|^2.
\end{eqnarray}
Another part of the static internal energy density is related to the normal state (when $\psi_\alpha=0$) in the absence of the MF (when $\nabla\times\mathbf{A}=0$); the normal energy density is denoted by $f_{n0}(\mathbf{r})=f_{n0}[n_p^\mathrm{n}(\mathbf{r}),n_n^\mathrm{n}(\mathbf{r})]$, and the energy of the normal state is
\begin{equation}\label{Fn}
F_n=\int d^3 \mathbf{r}\,f_{n0}.
\end{equation}
The SF excitation free energy density up to the next-to-next-to-leading order reads
\begin{eqnarray}
\nonumber
  &&f^{\mathrm{GL}}(\mathbf{r})=f_u[n_p,n_n]+f_{n0}[n_p^\mathrm{n},n_n^\mathrm{n}]+\frac{(\nabla\times\mathbf{A})^2}{8\pi}\\
  &&\nonumber +\frac{\hbar^2}{2m}|\hat{\mathbf{D}}\psi_p|^2+\frac{\hbar^2}{2m}|\nabla\psi_n|^2+\lambda_{np}(\nabla|\psi_p|)\cdot(\nabla|\psi_n|) \\
\nonumber
  &&-\frac{\hbar^2}{2m}n_{np}\left[\frac{1}{2\mathrm{i}} \left( \frac{\psi_p^*\hat{\mathbf{D}}\psi_p-c.c.}{|\psi_p|^2} - \frac{\psi_n^*\nabla\psi_n-c.c.}{|\psi_n|^2}  \right) \right]^2\\
\label{FreeEnergyGL}
  &&+\mu_{np}|\psi_p||\psi_n|(\nabla|\psi_p|)\cdot(\nabla|\psi_n|),
\end{eqnarray}
and the free energy $F^{\mathrm{GL}}$ of the system is
\begin{equation}\label{FGL}
F^{\mathrm{GL}}=\int d^3 \mathbf{r}\,f^{\mathrm{GL}}.
\end{equation}
\subsubsection{Comparison with earlier two-component models}
The free energy density in Eq. (\ref{FreeEnergyGL}) is equivalent to the earlier form of the GL model used in a general hydrodynamic formulation and in studies of some topological properties of the SC mixtures \cite{AndreevBashkin1976,VardanyanSedrakyan1981,AlparLangerSauls1984,MendellLindblom1991,Mendell1991,Prix2004,Babaev2004,GlampedakisAnderssonSamuelsson2011,KobyakovPethick2017,KobyakovEtal2017}; in these works the quantum pressure and the gradient-coupling are implicit.
The equations for wave functions with either only momentum-coupling, or only gradient-coupling were studied in  \cite{DrummondMelatos2017,AlfordGood2008,HaberSchmitt2017}.

In order to analyze relations between the different formulations it is convenient to expand the term corresponding to the momentum-coupling in the right-hand side of Eq. (\ref{FreeEnergyGL}).
Since the electromagnetic field is not essential here, it is ignored for a moment and Eq. (\ref{Eentr}) is used.
One obtains the following ten terms:
\begin{eqnarray}
  \nonumber&&\!\!\!\!\!\!\!\!\!\!\!\!\!\!\!\!\!\!\!\!\!\!\!\!\!\!\!\!\!\!\!\!\!\!\!\!-\frac{\hbar^2}{2m}n_{np}\left(\frac{1}{2\mathrm{i}}\nabla \ln\frac{\psi_p\psi_n^*}{\psi_p^*\psi_n}\right)^2\\
  \nonumber&&=\frac{\hbar^2n_{np}}{8mn_pn_n}\left(   -n_n|\nabla\psi_p|^2 -n_p|\nabla\psi_n|^2\right. \\
  \nonumber&&\quad+\frac{n_p}{n_n}\left[(\psi_n\nabla\psi_n^*)^2+c.c.\right] \\
  \nonumber&&\quad\quad+\frac{n_n}{n_p}\left[(\psi_p\nabla\psi_p^*)^2+c.c.\right]  \\
  \nonumber&&\quad\quad\quad-\left[(\psi_p\nabla\psi_p^*)(\psi_n\nabla\psi_n^*)+c.c. \right]\\
  \label{Uent2019}&&\quad\quad\quad\quad+\left.\left[(\psi_p\nabla\psi_p^*)(\psi_n^*\nabla\psi_n)+c.c.\right]\right).
\end{eqnarray}
The first six terms in the right-hand side of  Eq. (\ref{Uent2019}) represent a renormalization of the quantum kinetic energies (the terms ${\hbar^2}|\nabla\psi_\alpha|^2/{2m}$) of the fluids due to the momentum-coupling.
The presence of those terms is necessary for the Galilean invariance of the internal fluid energy.
The rest four terms of Eq. (\ref{Uent2019}) are equivalent to the ones used in equation (5) of \cite{AlparLangerSauls1984}.
The last four terms in Eq. (\ref{Uent2019}) were also used in equation (3) of \cite{AlfordGood2008}, however, the relative signs are different from the corresponding signs in equation (5) of \cite{AlparLangerSauls1984}.
This is not surprising because the calculations are focused on the momentum-coupling in \cite{AlparLangerSauls1984,DrummondMelatos2017}, and on the gradient-coupling in \cite{AlfordGood2008,HaberSchmitt2017}.
The "coefficients" by the terms of the form $(\psi_\alpha\nabla\psi_\alpha^*)(\psi_\beta\nabla\psi_\beta^*)$ in Eq. (\ref{Uent2019}) are seen to be non-trivial functionals of the SF densities, which is a consequence of the representation form via the wave function gradient.
However, once the whole expression is written in terms of the basic degrees of freedom as explained above, the perturbative structure of the excitation energy in the framework of the effective field theory becomes obvious.

A comparison of the basic functional in Eq. (\ref{FreeEnergyGL}) and of the functional displayed in equations (2) and (3) in Alford \& Good (2008) \cite{AlfordGood2008} reveals that the model suggested in \cite{AlfordGood2008} is a special case of the present model with $\lambda_{np}=0$ and $n_{np}=0$, and that the parameter $\sigma$ in equation (3) in \cite{AlfordGood2008} is irrelevant to the parameter $\varepsilon$ from \cite{LindblomMendell2000}, but it bears the role of the parameter $\mu_{np}$ of the present model.
The entrainment parameter $\varepsilon$ from \cite{LindblomMendell2000} is directly linked to the functional $n_{np}$ in Eq. (\ref{FreeEnergyGL}).

\section{Basic equations}
Equations for SF density structure and the phase result from minimization of the free energy of the system:
\begin{equation}\label{stationaryNLSE}
\partial(\delta F)/\partial\psi_\alpha^*=0,
\end{equation}
where $\delta F\equiv F^{\mathrm{GL}}-F_n$ \cite{LL1980}.
Using the perturbation energy given in Eq. (\ref{FreeEnergyGL}), by analogy with Eqs. (\ref{NLSE1}),(\ref{mutot1})-(\ref{Hid}),(\ref{NLSE1form})-(\ref{muntot}) I obtain
\begin{equation}\label{NLSE2form}
  \left[\mu^{\mathrm{tot}}_\alpha-\mathrm{i}\frac{1}{2n_\alpha}\nabla\left(n_\alpha\mathbf{v}_\alpha\right)\right]\psi_\alpha=0,
\end{equation}
with
\begin{eqnarray}
  \nonumber \mu_{\alpha}^{\rm tot}=\frac{P_\alpha^2}{2m}-\frac{\hbar^2}{2m}\frac{\nabla^2\sqrt{n_\alpha}}{\sqrt{n_\alpha}}-\frac{1}{2m}\frac{\partial n_{np}}{\partial n_{\alpha}}\left(\mathbf{P}_p-\mathbf{P}_n\right)^2\\
 \label{muAtot} -\frac{\lambda_{np}}{2\sqrt{n_\alpha}}\nabla^2 \sqrt{n_{\check{\alpha}}}-\frac{\mu_{np}}{4}\nabla^2 n_{\check{\alpha}}+\frac{\partial f_u}{\partial n_\alpha},
\end{eqnarray}
where the check implies that $\check{p}=n$ and $\check{n}=p$.
Equations for the wave functions can be cast to the form:
\begin{eqnarray}
\nonumber&& \!\!\!\!0=-\frac{\hbar^2}{2m}(\nabla-\mathrm{i}\frac{e}{\hbar c}\mathbf{A})^2\psi_p+(a_p+b_{p}|\psi_p|^2+b_{np}|\psi_n|^2)\psi_p\\
\nonumber&&  +\left[\mathrm{i}\frac{\hbar}{2mn_p}\nabla\left(n_{np}(\mathbf{P}_p-\mathbf{P}_n)\right)-\frac{1}{2m}\frac{\partial n_{np}}{\partial n_p}(\mathbf{P}_p-\mathbf{P}_n)^2\right]\psi_p\\
\nonumber&&-\frac{\lambda_{np}}{2}(\nabla^2 \sqrt{n_n})\frac{\psi_p}{\sqrt{n_p}}-\frac{\mu_{np}}{4}(\nabla^2 n_{n})\psi_p,\\
\label{GLp}\\
\nonumber&& \!\!\!\!0=-\frac{\hbar^2}{2m}\nabla^2\psi_n+(a_n+b_{n}|\psi_n|^2+b_{np}|\psi_p|^2)\psi_n\\
\nonumber&&+\left[\mathrm{i}\frac{\hbar}{2mn_n}\nabla\left(n_{np}(\mathbf{P}_n-\mathbf{P}_p)\right)-\frac{1}{2m}\frac{\partial n_{np}}{\partial n_n}(\mathbf{P}_p-\mathbf{P}_n)^2\right]\psi_n\\
\nonumber&&-\frac{\lambda_{np}}{2}(\nabla^2 \sqrt{n_p})\frac{\psi_n}{\sqrt{n_n}}-\frac{\mu_{np}}{4}(\nabla^2 n_{p})\psi_n,\\
\label{GLn}\\
&&\label{GLA}\nabla\times\nabla\times\mathbf{A}=\frac{4\pi}{c}\mathbf{j},
\end{eqnarray}
where the electric current $\mathbf{j}$ is given by
\begin{equation}\label{jelectric}
  \mathbf{j}=e\mathbf{J}_p,
\end{equation}
with $\mathbf{J}_p$ from Eq. (\ref{momentaDef}),(\ref{currdef}),(\ref{vel1}).
\subsection{Uniform mixtures}
\label{uniform}
Since the nuclear condensates are in the miscible phase \cite{BaymBethePethick1971}, the equilibrium SF densities $n_{\alpha0}$ are homogeneous functions of space except near the boundary of the SC body, where they change on the length scale of the coherence length.
Substituting into Eqs. (\ref{GLp})-(\ref{GLA}) the equilibrium solutions with $n_{\alpha}=n_{\alpha0}=\mathrm{const}$, $\phi_{\alpha}=0$ and $\mathbf{A}=0$, one obtains the following linear system of equations:
\begin{equation}\label{n0}
\left(
  \begin{array}{cc}
    b_p & b_{np} \\
    b_{np} & b_n \\
  \end{array}
\right)\left(
         \begin{array}{c}
           n_{p0} \\
           n_{n0} \\
         \end{array}
       \right)=\left(
         \begin{array}{c}
           -a_p \\
           -a_n \\
         \end{array}
       \right).
\end{equation}
The uniform SF densities are given by the solution to Eq. (\ref{n0}):
\begin{equation}\label{n0dir}
\left(
         \begin{array}{c}
           n_{p0} \\
           n_{n0} \\
         \end{array}
       \right)=\frac{1}{\Delta_b}\left(
  \begin{array}{cc}
    b_n & -b_{np} \\
    -b_{np} & b_p \\
  \end{array}
\right)\left(
         \begin{array}{c}
           -a_p \\
           -a_n \\
         \end{array}
       \right),
\end{equation}
where $\Delta_b=b_pb_n-b_{np}^2$.
Inside the nonSC phase, where the proton pairing is exponentially suppressed by the MF, the neutron SF density is
\begin{equation}\label{n1comp}
  n_{n0}^{\mathrm{single}}=-\frac{a_n}{b_n}.
\end{equation}
Below, we also shall need noninteracting SF proton density defined analogously:
\begin{equation}\label{np1comp}
  n_{p0}^{\mathrm{single}}=-\frac{a_p}{b_p}.
\end{equation}
\subsection{Parameterization of interactions}
Following \cite{KobyakovEtal2017} I assume that $n_{np}$ is a bilinear functional of the SF densities:
\begin{equation}\label{nnpbilinear}
n_{np}=\frac{n_{np0}}{n_{p0}n_{n0}}n_pn_n,
\end{equation}
where $n_{np0}=\mathrm{const}$.
For perspectives of numerical calculations I introduce dimensionless variables:
\begin{equation}\label{dimensionlessQ}
  \bar{x}=\frac{x}{\delta},\quad\bar{\psi}_\alpha=\frac{\psi_\alpha}{\sqrt{n_{\alpha0}}},\quad\bar{A}=\frac{A}{H_c\delta},\quad\bar{B}=\frac{B}{H_c}.
\end{equation}
The quantity $H_c$ is the thermodynamic MF that was first found by Haber and Schmitt (2017) \cite{HaberSchmitt2017} and is given in Eq. (\ref{Hc}).
In terms of the dimensionless functions $\bar{\psi}_\alpha$, Eq. (\ref{nnpbilinear}) takes on the form
\begin{equation}\label{nnpbilinearPsi}
n_{np}={n_{np0}}|\bar{\psi}_p|^2|\bar{\psi}_n|^2.
\end{equation}
In this case, the derivatives of $n_{np}$ with respect to $n_\alpha$ (which are physically dimensionless by their definition) can be written in the following form
\begin{equation}\label{VectorCoupling}
\frac{\partial {{n}_{np}}}{\partial {{n}_{p}}}=\gamma|\bar{\psi}_n|^2\quad \mathrm{and}\quad \frac{\partial {{n}_{np}}}{\partial {{n}_{n}}}=\gamma\eta|\bar{\psi}_p|^2,
\end{equation}
where the parameter $\gamma$, defined explicitly in Eq. (\ref{parameters}), represents a dimensionless strength of the momentum-coupling. The unit of length is equal to
\begin{equation}\label{deltaGL}
\delta=\sqrt{\frac{mc^2}{4\pi e^2n_{p0}}}.
\end{equation}
The quantity $\delta$ is reminiscent of the London penetration depth in a pure SC but it is a different quantity because $\delta$ includes a characteristic two-component quantity $n_{p0}$.
The quantity $\delta$ is neither the magnetic penetration depth in a SF-SC mixture with entrainment given in Eq. (\ref{delta}).
For convenience, I choose to define the unit of length, Eq. (\ref{deltaGL}), as the magnetic penetration depth in a mixture with artificially switched off entrainment.
Further below, we shall work only with the dimensionless versions of the quantities $x$, ${\psi}_\alpha$, $A$, $B$, and for simplicity, I shall drop the bars from now on.
According to the standard definition, the coherence length $\xi$ in a single-component proton SC is
\begin{equation}\label{xi2}
  \xi^2=-\frac{\hbar^2}{2ma_p}.
\end{equation}
Numerical values of the basic quantities in nuclear matter are not known exactly, but their magnitudes may be evaluated at typical proton number density $n_{p0}\sim0.008$ fm$^{-3}$, as $\delta\sim80$ fm and $\xi\sim30$ fm \cite{BaymPethickPines1969,Link2003}.
We shall use the following dimensionless parameters for characterization of the equilibrium state of the system:
\begin{eqnarray}
  \nonumber&&\kappa^2=\frac{\delta^2}{\xi^2},\quad\eta= \frac{n_{p0}}{n_{n0}},\quad \theta_1=\frac{a_n}{a_p},\quad \theta_2=\frac{b_{np}}{b_p},\quad \theta_3=\frac{b_{np}}{b_n},\\
  \label{parameters}&&\gamma=\frac{n_{np0}}{n_{p0}}, \quad \mu=\mu_{np}\frac{mn_{n0}}{2\hbar^2},\quad\lambda=\frac{\lambda_{np}m}{\hbar^2}.
\end{eqnarray}
The dimensionless parameters have simple physical meaning: $\kappa$ is a generalization of the GL parameter of a pure proton SC to a SC-SF mixture; $\theta_1$ is the ratio of the quadratic GL coefficients; $\theta_2$ and $\theta_3$ are the ratios of the interspecies quartic GL coefficient to the intraspecies quartic GL coefficients, correspondingly; $\eta$ is the ratio of the SF densities, which were discussed in Sec. \ref{TSNd}; $\gamma$ is the dimensionless strength of the vector coupling between the SFs in units of the proton SF density (the momentum-coupling); and $\lambda$ and $\mu$ are the dimensionless strengths of the vector coupling between the amplitudes of the SFs (the gradient-coupling) in the next-to-leading and in the next-to-next-to-leading orders, correspondingly.
A different definition of the parameters $\xi$ and $\kappa$ was used in Refs. \cite{AlfordGood2008} and \cite{HaberSchmitt2017}, however this is a matter of notation and does not change the physical conclusions.
The GL coefficients $a_\alpha$, $b_\alpha$ and $b_{np}$ may be evaluated following the approach of Gorkov \cite{Gorkov1959}.

Using the dimensionless parameterization I obtain some useful relations:
\begin{eqnarray}
  &&n_{p0}=n_{n0}^{\mathrm{single}}\frac{\theta_2}{\theta_1\theta_3}\frac{1-\theta_1\theta_3}{1-\theta_2\theta_3},\\
  \label{np0npsingle}&&n_{n0}=n_{n0}^{\mathrm{single}}\frac{1}{1+\eta\theta_3},\\
  &&\eta=\frac{\theta_2}{\theta_3}\frac{1-\theta_1\theta_3}{\theta_1-\theta_2},\\
  \label{theta1eta}&&\theta_1=\frac{\theta_2}{\theta_3}\frac{1+\eta\theta_3}{\eta+\theta_2},\\
  \label{nn0single}&&n_{n0}^{\mathrm{single}}=\frac{1+\eta\theta_3}{\eta+\theta_2}n_{p0}^{\mathrm{single}}.
\end{eqnarray}
Equation (\ref{theta1eta}) implies that a complete description requires one parameter less than shown in Eq. (\ref{parameters}): either $\theta_1$ or $\eta$ may be dropped without a loss of information.

It is important that two sets of the dimensionless parameters, $(\kappa,\theta_1,\theta,\lambda,\gamma,\mu)$ and $(\kappa,\eta,\theta,\lambda,\gamma,\mu)$, are equivalent.
The two sets are simplified and incomplete.
Not all of the next-to-next-to-leading order terms have been parameterized here.
In fact, I assumed $\mu_{\alpha\alpha}=0$, Eq. (\ref{muAA}), following the earlier literature \cite{AlfordGood2008}, however there is no obvious reason for such assumption and an extensive general map of the two-component model must include information on the cases with $\mu_{\alpha\alpha}\neq0$, however this goes beyond aims of the present paper.
In a phenomenological theory of nuclear matter, the proton quadratic GL coefficient may be obtained from the knowledge of the coherence length $\xi$, Eq. (\ref{xi2}), while the latter can be calculated from microscopic physics (for neutrons, the calculation is analogous); in this way, one finds $\theta_1$ and then evaluates $\eta$.
\subsection{Stability of uniform SF mixtures}
Stability of uniform mixtures with a SF counterflow was considered in \cite{KobyakovPethick2017}.
In a hydrostatic case, a solution is meaningful when
\begin{equation}\label{DensGEQ0}
n_{\alpha0}\geq0.
\end{equation}
Since the aim is to study the SC-SF interaction effects I assume that noninteracting condensates are characterized by nonnegative densities,
\begin{equation}\label{singleDensGEQ0}
n_{\alpha0}^{\mathrm{single}}\geq0.
\end{equation}
Both conditions are relevant within the problem of the order parameter structure at the normal-SC boundary in the MF.
The requirement $n_{n0}^{\mathrm{single}}\geq0$ implies that the neutron SF density is nonnegative in the depth of the pure neutron SF phase in the MF, while the requirement $n_{n0}\geq0$ implies that the SF neutron density is nonnegative within the SC-SF bulk.
From Eqs. (\ref{np0npsingle}), (\ref{DensGEQ0}) and (\ref{singleDensGEQ0}) I obtain the first condition of thermodynamic stability in the following form:
\begin{equation}\label{stability1}
  1+\eta\theta_3\geq0. \quad(\mathrm{thermodynamic}\; \mathrm{stability})
\end{equation}
The second condition results from Eqs. (\ref{nn0single}) and (\ref{singleDensGEQ0}):
\begin{equation}\label{stability2}
  \eta+\theta_2\geq0. \quad(\mathrm{thermodynamic}\; \mathrm{stability})
\end{equation}
Thermodynamically stable solutions may be dynamically unstable due to the fluctuations.
The long-wavelength dynamical stability condition has been found in \cite{KobyakovPethick2017} and reads
\begin{equation}
  \label{stability3}1-\theta_2\theta_3>0.\quad(\mathrm{long\;wavelength}\;\mathrm{dynamic}\; \mathrm{stability})
\end{equation}
\subsection{Special case $b_p=b_n$}
For the practical purposes, it is useful to consider a special case when $b_p=b_n\equiv b$ as has been shown by Alford \& Good  (2008) \cite{AlfordGood2008}.
In this case, the parameterization reduces to $\theta_2=\theta_3\equiv \theta$.
The stability conditions reduce to a form
\begin{equation}\label{stability11}
  {\eta+\theta}\geq0, \quad(\mathrm{thermodynamic}\; \mathrm{stability})
\end{equation}
and
\begin{equation}\label{stability21}
  1+\eta\theta\geq0. \quad(\mathrm{thermodynamic}\; \mathrm{stability})
\end{equation}
Figure 1 shows the inequalities in Eqs. (\ref{stability11}) and (\ref{stability21}) graphically.
The shaded (grey) area including the line $\eta=-\theta$ shows the region of the parameter space where physically meaningful solutions do not exist.
The solutions with parameters $(\eta,\theta)$, located on the black part of the curve $\eta=-1/\theta$, correspond to the case when the neutron SF density decays to zero in the bulk of the normal phase.
The white area at $\theta\leq0$ represents parameters that correspond to physically meaningful solutions $n_{\alpha0}>0$.
\begin{figure}
\includegraphics[width=3.5in]{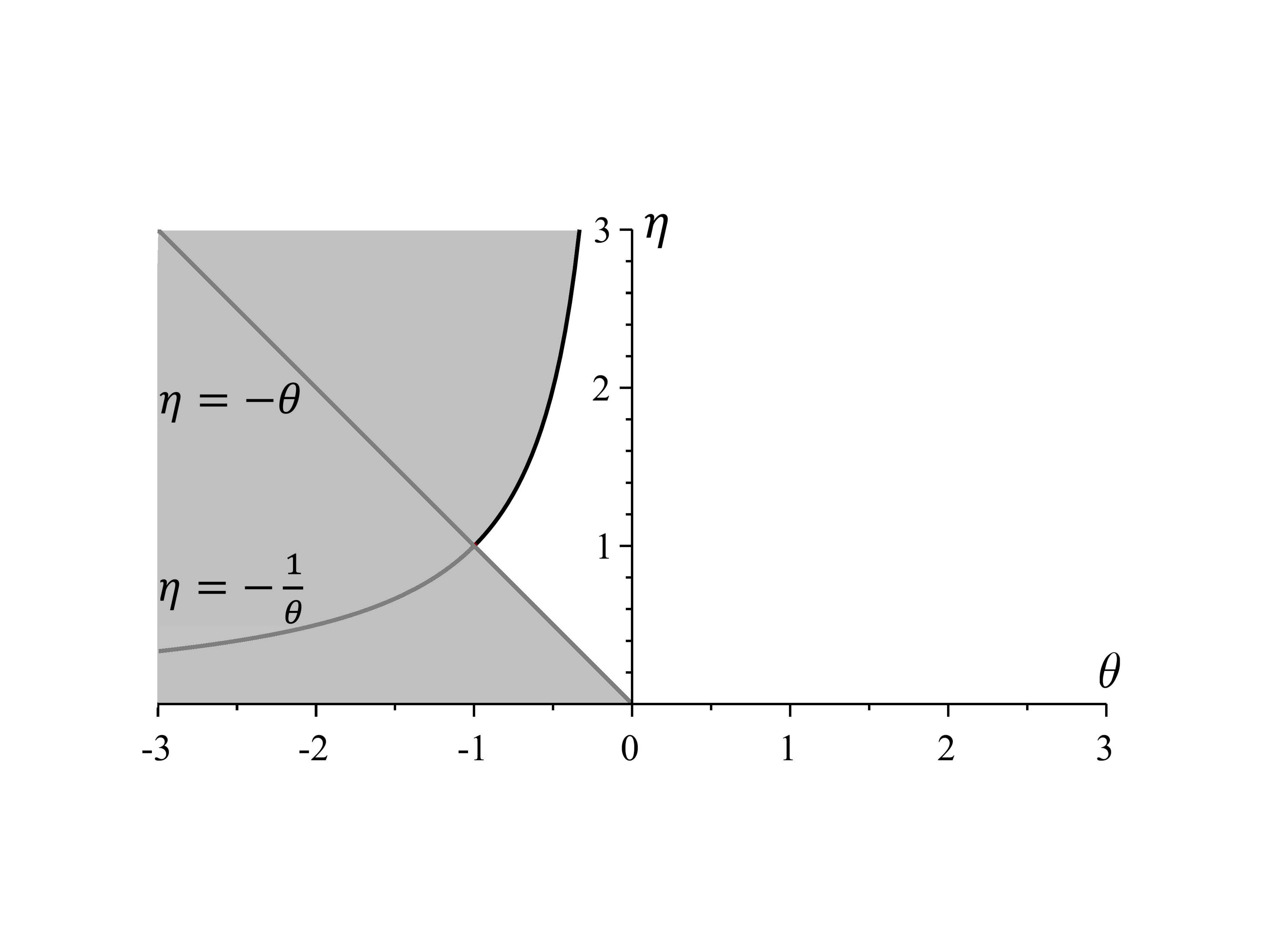}
\caption{Graphical representation of Eqs. (\ref{stability11}) and (\ref{stability21}) which provide parametric conditions for the thermodynamic stability of the uniform SF mixture with $b_p=b_n$ and $\theta\equiv b_{np}/b_p\leq0$ and in the absence of the matter flows and the SF density gradients.
White area at $\theta\leq0$ displays parameters corresponding to solutions with positive SF densities $n_{\alpha0}>0$.
The shaded (grey) area shows parameters that correspond to $n_{\alpha0}<0$ (in this region, physically meaningful solutions do not exist).
The black part of the curve labeled $\eta=-\frac{1}{\theta}$ corresponds to $n_{p0}>0$ and $n_{n0}^{\mathrm{single}}=0$.}
\end{figure}
\subsection{Planar normal-SC interface}
We shall calculate the structure of a planar domain wall separating two different phases.
The phase separation is a result of the MF screening by the SC currents at the edge of the SC body, and is essentially different from the phase separation due to the gravitational stratification.
Location of the surface is determined by combined effects of the gravitational stratification and the thermodynamic equilibration between the globally nonuniform stellar MF configuration and the order parameter amplitude (the latter is density dependent when calculated in the absence of the MF).
In this paper, it is assumed that a model with uniform total density is applicable to calculation of the SE in the NS matter.
This assumption is justified by noting that both the stratification length scale at which the total density varies and the length scale for local variations of the MF supported mostly by the electron currents in proto-NS are much longer than the domain wall width.
Therefore, to calculate the SE locally it is reasonable to assume that its geometric form is planar, and the total density and the MF are uniform.
Stability of the boundary at a fixed position in space is determined by equality of the quasiparticle chemical potentials at the both sides.
Equivalently, the quantities $a_\alpha$ in Eq. (\ref{fu}) are uniform.

Following \cite{LL1980} I consider a planar surface ($y-z$ plane) that separates the "normal" and SC phases, with the MF $\mathbf{H}$ along $z$ axis.
With this choice the wavefunctions become one-dimensional, $\psi_\alpha=\psi_\alpha(x)$, and the vector potential can be chosen as $\mathbf{A}=(0,A(x),0)$.
Since we consider a region in space where the wave function changes between zero and the bulk value (no rotation is considered), it is reasonable to introduce the MF $\mathbf{H}$ according to the definition
\begin{equation}\label{Hdef}
  \mathbf{H}=\nabla\times\mathbf{A}-4\pi\mathbf{M},\quad c\nabla\times\mathbf{M}=\mathbf{j}.
\end{equation}
Application of $\mathrm{curl}$ to Eq. (\ref{Hdef}) shows that the SC current is the source of the body magnetization $\mathbf{M}$, while the MF $H=\mathrm{const}$.
Deep in the "normal" region, sufficiently far from the interface, the magnetic induction is equal to the MF:  $\nabla\times\mathbf{A}=\mathbf{\hat{z}}H=\mathbf{\hat{z}}H_c$.
In the SC bulk, the magnetic induction $\mathbf{B}=\nabla\times\mathbf{A}$ decays according to the Londons' equation $\Delta\mathbf{B}=-\delta_L^2\mathbf{B}$ with the magnetic penetration depth
\begin{equation}\label{delta}
  \delta_L=\sqrt{\frac{mc^2}{4\pi e^2(n_{p0}-n_{np0})}}.
\end{equation}
The quantity $n_{np0}$ is equal to the entrainment density $n_{np}$ evaluated with the uniform SF densities $n_{\alpha0}$ given in Eq. (\ref{n0dir}):
\begin{equation}\label{nnp0Def}
  n_{np0}\equiv n_{np}[n_p=n_{p0},n_n=n_{n0}].
\end{equation}
The boundary conditions for the dimensionless functions (corresponding to the "normal" and SC phases at $x\rightarrow-\infty$ and $x\rightarrow+\infty$) are:
\begin{eqnarray}
\nonumber&&{{\psi }_{p}}=0,\quad{{\psi }_{n}}=\sqrt{1+\eta\theta_3}, \quad {A}'=1 \quad {\mathrm{at}} \quad x=-\infty, \\
\label{infM}\\
\nonumber&&{{\psi }_{p}}=1, \quad {{\psi }_{n}}=1, \quad {A}'=0 \quad \mathrm{{at}} \quad x=+\infty.\\
\label{infP}
\end{eqnarray}
In the normal phase, the asymptotic MF is given by Eq. (\ref{Hc}) because that keeps the neutron quadratic GL coefficient constant across the boundary and the boundary is stable.
The dimensionless equations describing the structure of the normal-SC interface are
\begin{eqnarray}
\label{basic1}
&&\frac{{{d}^{2}}}{{d}{{{x}}^{2}}}{{\psi }_{p}}=-\mu \frac{{{d}^{2}}\left( {{\left| {{\psi }_{n}} \right|}^{2}} \right)}{{d}{{x}^{2}}}{{\psi }_{p}}-\frac{\lambda}{\sqrt{\eta}}\frac{{{d}^{2}|\psi_n|}}{{d}{{{x}}^{2}}}\\
&&\nonumber+\frac{{{\varkappa }^{2}}}{2}\frac{1-{{\theta }_{2}}{{\theta }_{3}}}{1+{{\theta }_{2}}/\eta }{{A}^{2}}\left( 1-\frac{\partial {{n}_{np}}}{\partial {{n}_{p}}} \right){{\psi }_{p}}\\
\nonumber
&&-{{\varkappa }^{2}}\frac{1-{{\left| {{\psi }_{p}} \right|}^{2}}+\left( 1-{{\left| {{\psi }_{n}} \right|}^{2}} \right){{\theta }_{2}}/\eta }{1+{{\theta }_{2}}/\eta }{{\psi }_{p}},\\
&&\frac{{{{d}}^{2}}}{{d}{{{x}}^{2}}}{{\psi }_{n}}=-\eta \mu \frac{{{d}^{2}}\left( {{\left| {{\psi }_{p}} \right|}^{2}} \right)}{d{{x}^{2}}}{{\psi }_{n}}-{\lambda}{\sqrt{\eta}}\frac{{{d}^{2}|\psi_p|}}{{d}{{{x}}^{2}}}\\
\nonumber&&-{{\varkappa }^{2}}\left( {{\theta }_{1}}-\frac{\theta_2}{\theta_3}\frac{{{\left| {{\psi }_{n}} \right|}^{2}}+{{\left| {{\psi }_{p}} \right|}^{2}}\eta\theta_3}{\eta +{{\theta }_{2}}}+\frac{\eta }{2}\frac{1-{{\theta }_{2}}{{\theta }_{3}}}{\eta +{{\theta }_{2}}}\frac{\partial {{n}_{np}}}{\partial {{n}_{n}}}{{A}^{2}} \right){{\psi }_{n}},\\
\label{basic3}
&&\frac{{{{d}}^{2}}}{{d}{{x}^{2}}}A=\left( {{\left| {{\psi }_{p}} \right|}^{2}}-\frac{{{n}_{np}}\left[ {{n}_{p}},{{n}_{n}} \right]}{{{n}_{p0}}} \right)A,
\end{eqnarray}
Equations (\ref{basic1})-(\ref{basic3}) with boundary conditions given in Eqs. (\ref{infM}) and (\ref{infP}) represent the first part of the numerical problem that is in focus in this paper.
In addition, for a numerical solution we shall need the SE integral, which will be discussed after introducing the critical MFs.

\section{Two types of superconductivity}
According to the thermodynamic interpretation, sign of the SE of the order parameter immersed in the MF directly defines whether superconductivity is type-I (positive SE) or type-II (negative SE).
Within the traditional classification \cite{LL1980} one does not mark out the case of vanishing SE as a separate type of superconductivity, and therefore it is reasonable to consider \emph{two} types of superconductivity supplemented by the special case of vanishing SE.
In the single-component case, the criterion of superconductivity type is given by sign of the expression $(\kappa-\frac{1}{\sqrt{2}})$: negative for type-I and positive for type-II.
The superconductivity type of a two-component SF-SC mixture is determined by more a complicated expression which includes the interaction parameters.
As I show below in this section, it is possible to analytically predict what type of superconductivity a material with given parameters has, and for this we shall need the notion of the critical MFs.
\subsection{Thermodynamic critical MF $H_c$}
Thermodynamic critical MF $H_c$ in a SF-SC system has been found in by Haber and Schmitt (2017) \cite{HaberSchmitt2017}.
This is a non-trivial problem because it is unclear immediately what is the behaviour of the neutron SF condensate.
The field $H_c$ points to the amount of the magnetic energy which is necessary to destroy the SC Cooper pairs by the MF.
Before and after the phase transition, the total energy is conserved.
A natural feature of a two-component system is that there are in principle two options for the final state that conserves the total free energy: either to change the SF neutron quadratic GL coefficient keeping the SF neutron density fixed, or to keep constant SF neutron quadratic GL coefficient and to let the SF neutron density change.
The scenarios are characterized by uniform MFs with different magnitudes.
The MF associated with the process when the neutron chemical potential is constant across the phase transition is lower than the MF associated with the process when the neutron SF density is constant across the phase transition.
The phase transition is expected to occur at a lower critical MF.
This consideration confirms the result found by Haber \& Schmitt \cite{HaberSchmitt2017} for the thermodynamic critical MF.

In a two-component system, the total free energy is given by Eq. (\ref{FGL}).
It is convenient to introduce the GL free energy $F^{\mathrm{GL}}_0$ of a uniform state with SF densities $n_p$ and $n_n$:
\begin{equation}\label{FGL0}
  F^{\mathrm{GL}}_0[n_p,n_n]=\int d^3\mathbf{r}\: f_u[n_p,n_n]+f_{n0},
\end{equation}
where $f_{n0}$ and $f_u$ are given in Eqs. (\ref{Fn}) and (\ref{fu}), and the SF densities $n_{\alpha}=n_{\alpha0}$ are given by Eq. (\ref{n0dir}).
The thermodynamic critical MF $H_c$ is obtained from the following expression
\begin{equation}\label{HcDefinition}
  -{H_c^2}/8\pi=F^{\mathrm{GL}}_0[n_{p0},n_{n0}]-F^{\mathrm{GL}}_0[0,n_{n0}^{\mathrm{single}}].
\end{equation}
Substituting the uniform SF densities from Eq. (\ref{n0dir}) into Eq. (\ref{HcDefinition}) one finds the thermodynamic critical MF:
\begin{equation}\label{Hc}
  H_c=n_{p0}\sqrt{4\pi b_p\left(1-\frac{b_{np}^2}{b_pb_n}\right)}.
\end{equation}
In the single-component limit (when $a_n=b_n=b_{np}=0$), the result in Eq. (\ref{Hc}) reduces to the well-known expression $\sqrt{4\pi a_p^2/b_p}\equiv n_{p0}^{\mathrm{single}}\sqrt{4\pi b_p} $ \cite{LL1980}.
\subsection{The upper critical MF $H_{c2}$}
For simplicity, I assume here $\lambda=0$.
The upper critical MF $H_{c2}$ is the smallest field consistent with motion of a quantum-mechanical particle at a given energy level.
For the field strength slightly below $H_{c2}$, the proton order parameter is small enough and $|\psi_p|^2\rightarrow0$, while the neutron order parameter is uniform with the SF density given by Eq. (\ref{n1comp}).
In this case, the non-vanishing contribution from the momentum-coupling is readily calculated using Eq. (\ref{VectorCoupling}),
\begin{equation}\label{eqMomCoupHc2}
  \frac{\partial n_{np}}{\partial n_p}=\gamma\frac{n_{n0}^{\mathrm{single}}}{n_{n0}}=\gamma(1+\eta\theta_3),
\end{equation}
and the equations of motion, Eqs. (\ref{GLp}) - (\ref{GLA}), with dimensionless functions $\psi_\alpha$ and $\mathbf{\mathrm{A}}$ as discussed above, become
\begin{eqnarray}
&&\nonumber
\frac{\hbar^2}{2m}\left(\nabla-\mathrm{i}\frac{e}{\hbar c}H_c\delta\mathbf{A}\sqrt{1-\gamma(1+\eta\theta_3)}\right)^2\psi_p\\
&&\label{equationHc21}\qquad=(a_p+b_{np}n_{n0}|\psi_n|^2)\psi_p, \\
&&\label{equationHc22}  |\psi_n|^2=\frac{n_{n0}^{\mathrm{single}}}{n_{n0}},\\
&&\label{equationHc23} \nabla\times\nabla\times\mathbf{A}=0.
\end{eqnarray}
It is easy to observe that the $\mu_{np}$-coupling terms vanish due to the assumption $|\psi_p|^2\rightarrow0$ and because the neutron density is uniform.
Equation (\ref{equationHc21}) is just the Schr$\ddot{\mathrm{o}}$dinger equation for a particle with mass $m$ and charge $e\sqrt{1-\gamma(1+\theta\eta)}$ in the MF.
The quantity $(a_p+b_{np}n_{n0}^{\mathrm{single}})$ is a stationary energy level.
Analogously to the case considered in detail in \cite{LL1980}, I immediately find that the critical field is given by the following expression
\begin{equation}\label{Hc2dim}
  H_{c2}=-\frac{2mca_p}{e\hbar}\frac{\eta}{\sqrt{1-\gamma(1+\eta\theta_3)}}\frac{1-\theta_2\theta_3}{\eta+\theta_2}.
\end{equation}
When the momentum-coupling is absent $n_{np}=0$ (or $\gamma=0$), Eq. (\ref{Hc2dim}) reduces to equation (13) of \cite{SinhaSedrakian2015} and equation (24) of \cite{HaberSchmitt2017}.
Furthermore, when the density-coupling and momentum-coupling are absent one has $b_{np}=0$ and $n_{np}=0$ (or $\theta=\gamma=0$), so Eq. (\ref{Hc2dim}) reduces to the single-component result \cite{LL1980}.
Finally, using
\begin{equation}\label{kappaDef}
\kappa^2=\frac{-a_p}{2\pi n_{p0}}\left(\frac{mc}{e\hbar}\right)^2,
\end{equation}
I obtain the upper critical field cast in the form
\begin{equation}\label{Hc2dimless}
  H_{c2}=H_c\kappa\sqrt{\frac{2\eta \left( 1-{{\theta }_{2}}{{\theta }_{3}} \right)}{\left( \eta +{{\theta }_{2}} \right)\left( 1-\gamma \left( 1+\eta {{\theta }_{3}} \right) \right)}},\quad(\lambda=0).
\end{equation}
\subsection{The critical GL parameter $\kappa_c$}
It is convenient to characterize SC materials with the help of the GL parameter $\kappa$ (generalization for two component systems was introduced in Eq. (\ref{parameters})), which is unambiguously connected to the SE $\alpha_{ns}$, since $\alpha_{ns}=\alpha_{ns}(\kappa)$ is a monotonic and decreasing function of $\kappa$.
The mathematical definition of $\alpha_{ns}$ is given in Eq. (\ref{alphaNS}).
The critical GL parameter $\kappa_c$ is defined by the following condition:
\begin{equation}\label{kappaCdef}
  \alpha_{ns}(\kappa=\kappa_c)\equiv0.
\end{equation}
Likewise, the critical fields are some functions of the GL parameter: $H_c=H_c(\kappa)$ and $H_{c2}=H_{c2}(\kappa)$. They represent outlooks upon the system from a perspective of either the first order (for $H_c$) or the second order phase transitions (for $H_{c2}$).
$\kappa_c$ represents a special parameter point, where the system simultaneously possess properties of the both SC regimes, and therefore in a system with $\kappa=\kappa_c$ the following equality holds:
\begin{equation}\label{betweenTypes}
 H_c(\kappa=\kappa_c)\equiv H_{c2}(\kappa=\kappa_c).
\end{equation}
Equations (\ref{Hc}), (\ref{Hc2dim}) and (\ref{betweenTypes}) represent the analytical solution to $\kappa_c$ as function of the coupling parameters of the SC-SF mixture.
I find the critical GL parameter from Eqs. (\ref{Hc}), (\ref{Hc2dimless}) and (\ref{betweenTypes}):
\begin{equation}\label{kappaCdens}
  \kappa_c=\sqrt{\frac{ 1 }{ 2 }\left(1 +\frac{{{\theta }_{2}}}{\eta} \right)\frac{ 1-\gamma \left( 1+\eta {{\theta }_{3}} \right) }{  1-{{\theta }_{2}}{{\theta }_{3}} }},\quad(\lambda=0).
\end{equation}
\subsection{Effect of the density-coupling on $\kappa_c$}
Does the bare density-coupling alone increase or decrease the GL critical parameter $\kappa_c$?
This question may be addressed by extracting the leading-order contribution in terms of the density-coupling parameter $\theta$, where, for simplicity and for purposes of the comparison with the earlier results we have assumed $b_p=b_n$.
Expansion at small $(\theta/\eta)$ with $\gamma=0$ yields
\begin{equation}\label{kappaCsmall}
\kappa_c\approx(1/\sqrt{2})(1+\frac{1}{2}(\theta/\eta)+O((\theta/\eta)^2)).
\end{equation}
Equation (\ref{kappaCsmall}) shows that effect of the bare density-coupling (which is negative in the nuclear matter, $\theta<0$) is to \emph{decrease} the critical GL parameter.
This qualitative conclusion and the quantitative result found in Eq. (\ref{kappaCdens}) provide a satisfactory fit to the numerical results given in Table I, which are obtained from the direct numerical solution of the basic equations of the model.
\subsection{Earlier work on $\kappa_c$}
It is instructive to note that the conclusion regarding the effect of the density-coupling on $\kappa_c$ reached by Alford \& Good (2008) using the asymptotic interactions between fluxtubes is equivalent to the conclusion reached in this paper.
Figure 4a in \cite{AlfordGood2008} shows that the dependence $\kappa_c(b_{np})$ is quadratic, however, this is a matter of definition of the GL parameter; if the calculations in \cite{AlfordGood2008} were presented in terms of the GL parameter introduced here, they would display a linear dependence.
Note that the GL parameter introduced by Alford \& Good (2008) differs from the GL parameter $\kappa$ introduced in this paper because the coherence length is defined in two alternative ways, equation (15) in Ref. \cite{AlfordGood2008} and Eq. (\ref{xi2}) in this paper.
For convenience, the GL parameter from Alford \& Good (2008) is denoted here as $\kappa_A$, and in terms of parameters of the present model it reads
\begin{equation}\label{kappaA}
  \kappa_A^2=\kappa^2\frac{\eta}{\eta+\theta_2}.
\end{equation}
Using Eq. (\ref{Hc2dimless}) one obtains
\begin{equation}\label{HcwithkA}
  H_{c2}=H_{c}\kappa_{A}\sqrt{\frac{2\left( 1-{{\theta }_{2}}{{\theta }_{3}} \right)}{1-\gamma \left( 1+\eta {{\theta }_{3}} \right)}}.
\end{equation}
It is easy to see that in case of bare denisity-coupling ($\gamma=0$) a small parameter expansion of the result for $\kappa_c$ obtained from the condition given in Eq. (\ref{betweenTypes}) with Eq. (\ref{HcwithkA}) generates a quadratic leading-order dependence on the density-coupling parameter $b_{np}$ represented by ${{\theta }_{2}}$ and ${{\theta }_{3}}$.
This implies that the results on the bare density-coupling here and in Ref. \cite{AlfordGood2008} are equivalent, however they are given in terms of different set of dimensionless parameters.

\section{Surface energy at a planar boundary between SC and non-SC phases}
\label{SurfaceTension}
In case when the MF induced by free currents $\mathbf{H}$ is present, the quantity $\tilde{f}$ is related to the free energy density of the system $f$ as follows:
\begin{equation}\label{GFE}
  \tilde{f}=f-\frac{\mathbf{H}\cdot\mathbf{B}}{4\pi}.
\end{equation}
The SE $\alpha_{ns}$ is the excess energy associated with interface of the SC phase immersed into the non-SC phase (where the matter is not superconducting but it may be superfluid) in the MF:
\begin{equation}\label{alphaNS}
  \alpha_{ns}=\int_{-\infty}^{+\infty}dx\quad \tilde{f}_{\mathrm{GL}}-\tilde{f}_n,
\end{equation}
with ${f}_{\mathrm{GL}}$ given by Eq. (\ref{FreeEnergyGL}) and $\tilde{f}_{\mathrm{GL}}=f_{\mathrm{GL}}-{H_cB}/{4\pi}.$
The free energy density of the non-SC phase in the MF is
\begin{equation}\label{fn}
  {f}_n=f_{n0}+\frac{H_c^2}{8\pi}+a_nn_{n0}^{\mathrm{single}}+\frac{b_n}{2}(n_{n0}^{\mathrm{single}})^2.
\end{equation}
Calculation of Eq. (\ref{alphaNS}) yields
\begin{widetext}
\begin{eqnarray}
  \nonumber&& {{\alpha }_{ns}}=\frac{H_{c}^{2}\delta }{8\pi}\frac{1}{ \left( 1-{{\theta }_{2}}{{\theta }_{3}} \right)}\underset{-\infty }{\overset{+\infty }{ \int }}\,dx~~~{{\psi }_{p}}^{4}+\frac{{{\theta }_{2}}}{{{\eta }^{2}{\theta }_{3}}}\left[ {{{\left( 1+\eta {{\theta }_{3}} \right)}^{2}}}+{{\psi }_{n}}^{4}+2\eta{{{\theta }_{3}}}{{\psi }_{p}}^{2}{{\psi }_{n}}^{2}-2\left( 1+\eta{{\theta }_{3}} \right){{\psi }_{n}}^{2} \right]\\
  \label{ST}&&+\frac{2}{{{\kappa }^{2}}}\left( 1+\frac{{{\theta }_{2}}}{\eta } \right)\left[ \psi {{_{p}^{'}}^{2}}+\frac{\psi {{_{n}^{'}}^{2}}}{\eta }-\frac{\lambda}{\sqrt{\eta}}\psi _{p}^{'}\psi _{n}^{'}-4\mu {{\psi }_{p}}{{\psi }_{n}}\psi _{p}^{'}\psi _{n}^{'} -\kappa^2{{\psi }_{p}}^{2}\right] +\left( 1-{{\theta }_{2}}{{\theta }_{3}} \right)\left[ \left( 1-\gamma {{\psi }_{n}}^{2} \right){{\psi }_{p}}^{2}{{A}^{2}}+{{\left( 1-{A}' \right)}^{2}} \right].
\end{eqnarray}
\end{widetext}
As expected, the integrand in Eq. (\ref{ST}) smoothly tends to zero at $x\rightarrow\pm\infty$.
The single-component expression is easily recognized within the right hand side of Eq. (\ref{ST}) when the interactions are turned off ($\theta_2=\theta_3=\gamma=\lambda=\mu=0$ and $\theta_2/\theta_3\rightarrow1$).

For numerical evaluation of the integral in Eq. (\ref{ST}), it is convenient to eliminate the gradient terms by using the equations of structure of the boundary, Eqs. (\ref{basic1})-(\ref{basic3}), and I obtain (note that the functions $\psi_\alpha$ and $A$ are dimensionless, as has been described above):
\begin{widetext}
\begin{eqnarray}
  \nonumber&&{{\alpha }_{ns}}=\frac{H_{c}^{2}\delta }{8\pi }\frac{1}{\left( 1-{{\theta }_{2}}{{\theta }_{3}} \right)}\underset{-\infty }{\overset{+\infty }{\mathop \int }}\,dx~~~~\psi _{p}^{2}\left[ 6\frac{ 1+{{{\theta }_{2}}}/{\eta }}{{{\kappa }^{2}}}\left(\mu \frac{{{d}^{2}}\left( {{\psi }_{n}}^{2} \right)}{d{{x}^{2}}}+\frac{\lambda}{\psi_p\sqrt{\eta}}\frac{{{d}^{2}} {{\psi }_{n}}}{d{{x}^{2}}}\right)-2\frac{{{\theta }_{2}}}{\eta }{{\psi }_{n}}^{2} \right]+\frac{{{\theta }_{2}}}{{{\eta }^{2}}{{\theta }_{3}}}\left[ {{\left( 1+\eta {{\theta }_{3}} \right)}^{2}}-{{\psi }_{n}}^{4} \right] \\
  \label{STnum}&&+\left( 1-{{\theta }_{2}}{{\theta }_{3}} \right)\left[ {{\left( 1-{A}' \right)}^{2}}+\gamma \psi _{p}^{2}{{\psi }_{n}}^{2}{{A}^{2}} \right]-\psi _{p}^{4}.
\end{eqnarray}
\end{widetext}

\section{Numerical results}
\label{numerical}
\subsection{Plan of calculations}
In order to obtain profiles of the functions $\psi_\alpha(x)$ and $A(x)$ at the normal-SC interface I solve Eqs. (\ref{basic1})-(\ref{basic3}).
First, I consider the single-component limit of Eqs. (\ref{basic1})-(\ref{basic3}) with $\kappa=1/\sqrt{2}$ on various spatial grids and we choose appropriate one.
The spatial derivative is approximated by three-point finite differences.
Solutions are found by the steepest descent method, starting from physically motivated trial functions.
The initial trial functions are: $\psi_p=(1+\tanh(\kappa/\sqrt{2}x))/2$, the function $\psi_n$ is constructed from the step function and $A$ is taken as $x$ multiplied by the step function.
I find that spatial domain $-40\leq x\leq40$ is suitable for the simulations because it allows for sufficiently uniform solutions away from the interface.
The solution is found by propagation of the system in imaginary time with the timestep $10^{-4}$.
The error is evaluated by substituting the numerical solution to the right-hand side of Eqs. (\ref{basic1})-(\ref{basic3}) and evaluating the deviation from zero of the difference between the left and right hand sides.
The numerical error of the solution can be systematically decreased by refining the grid.

The numerical value of $\alpha_{ns}$ is given by terms to the right from the factor of ${H_{c}^{2}\delta }/{8\pi }$ in the right hand side of Eq. (\ref{STnum}).
In the single-component limit, this quantity becomes smaller than 0.01 at grid size equal to or larger than 4096, with precision of the input solution below $4.01\times10^{-5}$.
It is therefore concluded that uncertainty with magnitude below $4.01\times10^{-5}$ of solutions $\psi_\alpha$ and $A$ for the boundary value problem generates numerical error with magnitude 0.01 in the final functional $\alpha_{ns}[\psi_p,\psi_n,A]$.
When the interactions are switched on the precision of calculation of the numerical value of $\alpha_{ns}$ is assumed to be equal to 0.01, by analogy with the single-component case.

Next, effects of the coupling parameters ($\theta$, $\gamma$, $\lambda$, $\mu$) are studied.
For each set of coupling parameters I sweep $\kappa$ seeking $\kappa_c$ according to the definition given in Eq. (\ref{kappaCdef}).
In this paper we study the equations numerically at $\eta=0.05$, but this parameter might be different; with this choice, it is convenient to compare our results with the earlier calculations performed in \cite{AlfordGood2008}.
Numerical data is compared with the analytical prediction, Eq. (\ref{kappaCdens}) in Table I.
Results for the gradient-coupling are not shown because this coupling leads to variation of the critical GL-parameter only in the fourth significant digit.
\subsection{Structure of the interface}
A characteristic profile of the interface is shown in Fig. 2.
One observes that on larger scales the interface profile looks very similar to the single-component case except that now there is also a neutron SF component.
Figure 3 shows characteristic features caused by various SF couplings obtained by solution of Eqs. (\ref{basic1})-(\ref{basic3}) with $\kappa=\kappa_c$ with $\kappa_c$ computed via Eq. (\ref{kappaCdens}) for each set of the parameters.
In all cases, the material is ideal diamagnetic, as expected.
The local minimum of the MF is a typical feature of the numerical solution found in both single and two component cases.
The density-coupling leads to variation of the boundary conditions for the neutron SF density.
The momentum-coupling affects the neutron wave function only in vicinity of the interface causing a local depletion of the SF neutron density.
The most exotic effect for the SF neutron density is produced by the gradient-coupling, which leads to appearance of a frozen localized wave of the SF neutron density, however this type of coupling has almost no effect on the SE integral.

The lower panel in Fig. 3 displays solutions for two sets of the parameters, $(\kappa,\eta,\theta,\lambda,\gamma,\mu)=(\kappa_c,0.05,0,0.5,0,0)$  in black lines and $(\kappa,\eta,\theta,\lambda,\gamma,\mu)=(\kappa_c,0.05,0,0,0,0.5)$ in green (gray) line only for the neutron SF density.
This panel shows effects of the gradient-coupling provided either by $\lambda_{np}$ alone, or by $\mu_{np}$ alone.
The functions $\psi_p^2$ and $B$ are not shown for the second set of parameters because those functions are essentially the same as in the case for the first set.
In accordance with the power counting scheme, the $\lambda_{np}$-coupling provides more significant effects on the SF neutron density than the $\mu_{np}$-coupling.

\begin{figure}
\includegraphics[width=3.5in]{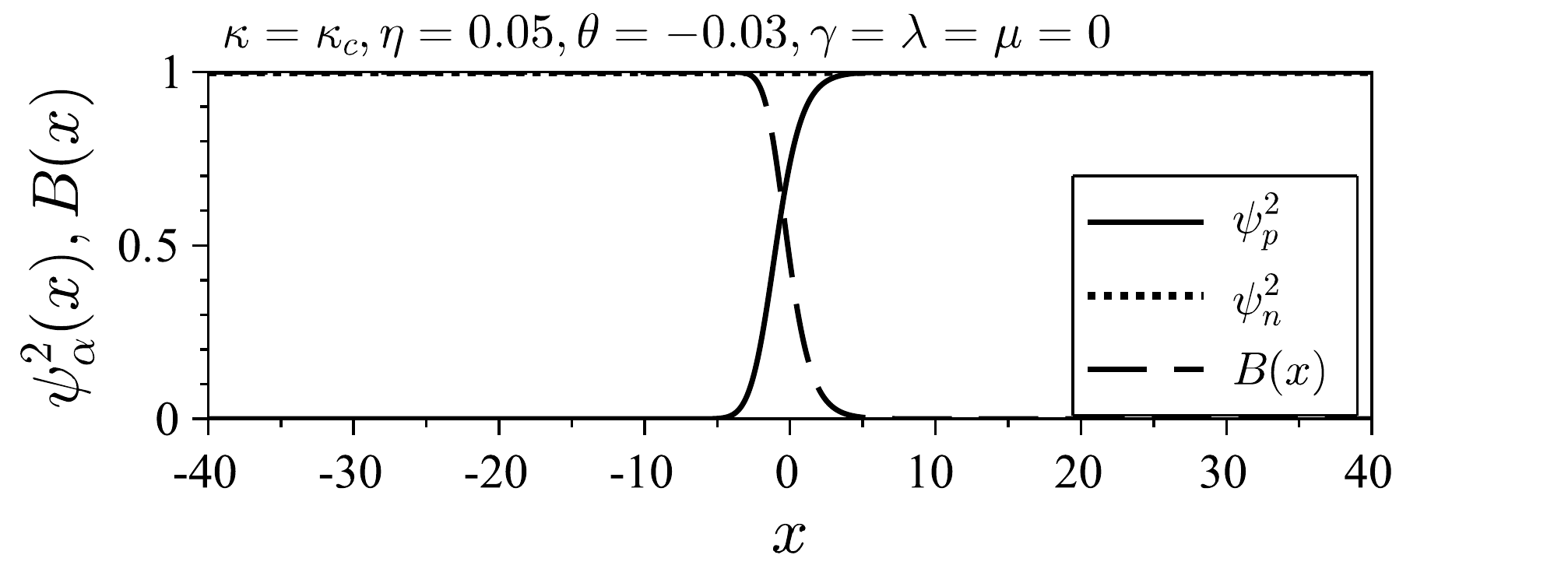}
\caption{Typical profiles of the SF densities and the MF at the normal-SC interface.
Equations (\ref{ntot}), (\ref{dimensionlessQ}) and (\ref{Hc}) provide dimensional factors.}
\end{figure}

\begin{figure}
\includegraphics[width=3.5in]{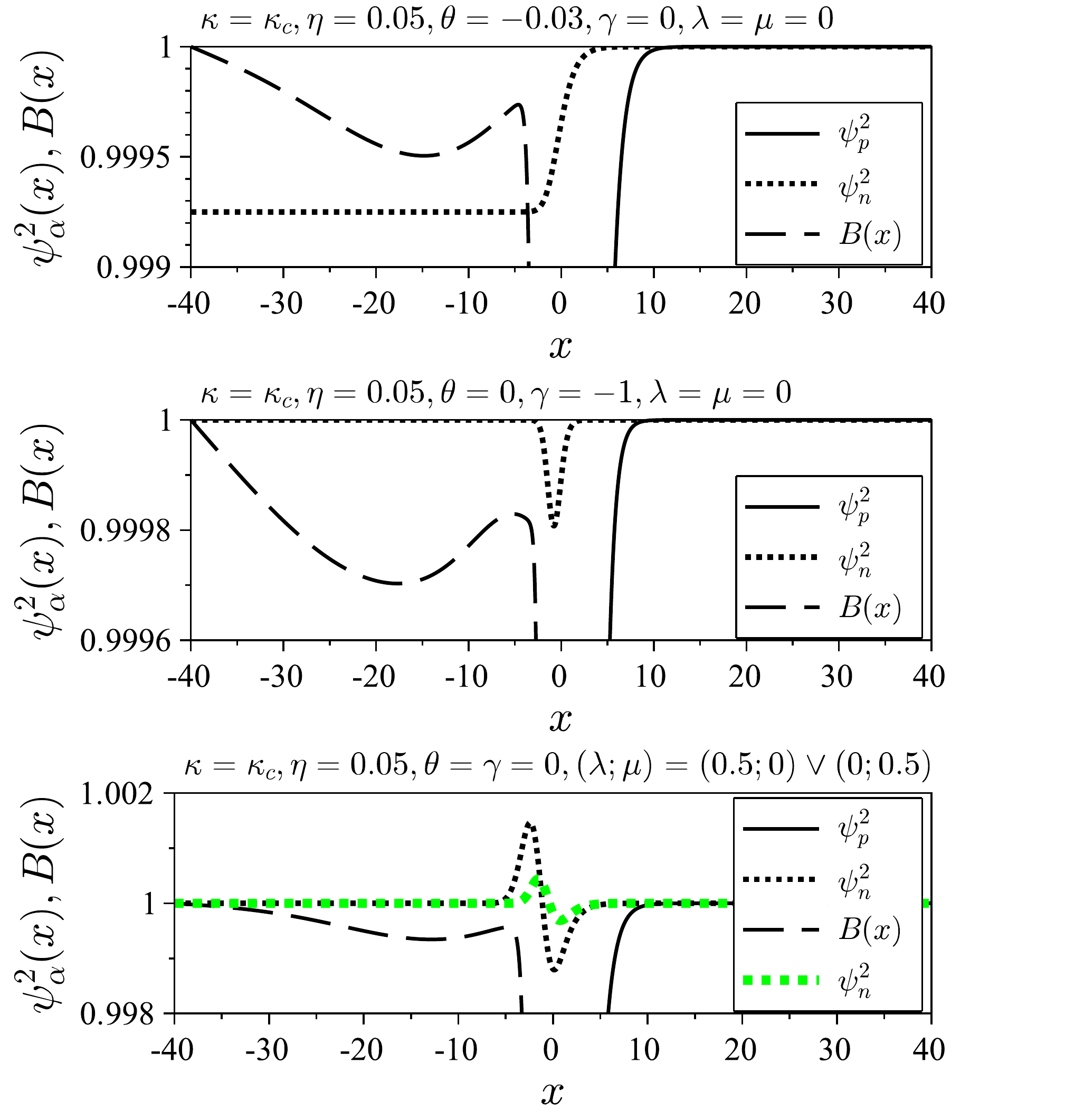}
\caption{Magnified view on characteristic features that arise due to various couplings between the SF and the SC.}
\end{figure}

\begin{table}
\caption{Analytical prediction and numerical data on the critical GL-parameter $\kappa_c$, Eq. (\ref{kappaCdef}), for various magnitudes of the bare density-coupling and bare momentum-coupling at $\eta=0.05$.}
\label{T1}
\begin{center}
\begin{tabular}{|c|c|c|c|}
  \hline
  $\theta$&$\gamma$&$\kappa_{c}\;(\mathrm{Eq}.\:(\ref{kappaCdens}))$&$\kappa_{c}$\\ \hline

0 & 0  & $1/\sqrt{2}$ & $1/\sqrt{2}$\\
-0.005  & 0& 0.6708 & 0.6714\\
-0.010  & 0& 0.6325 & 0.6326\\
-0.015  & 0& 0.5917 & 0.5917\\
-0.020  & 0& 0.5478 & 0.5479\\
-0.025  & 0& 0.5002 & 0.5003\\
-0.030  & 0& 0.4474 & 0.4475\\
0 & -0.2 & 0.7746 & 0.7748\\
0 & -0.4 & 0.8367 & 0.8371\\
0 & -0.6 & 0.8944 & 0.8948\\
0 & -0.8 & 0.9487 & 0.9490\\
0 & -1.0 & 1.000 & 1.001\\

 \hline
\end{tabular}
\end{center}
\end{table}

Table I provides a basic check of the analytical prediction given in Eq. (\ref{kappaCdens}) for bare density-coupling and momentum-coupling.
As Table I shows for the bare couplings near the origin for $\eta=0.05$, the density-coupling provides an additional negative SE at the normal-SC interface, leading to a decrease of the SE, thus pushing the system towards the type-II superconductivity regime.
The effect of the momentum-coupling $\gamma$ turns out to be the opposite: it provides an additional positive SE.
The data on the gradient-coupling is not shown because its effect on the SE is negligible.
In principle, the superconductivity type may become I even with $\kappa\sim3$ if $\kappa_c$ becomes larger than $\sim3$, which might be possible for parameter sets with $\theta\lesssim-1$ and $\eta\gtrsim1$.

\section{Conclusions}
\label{Conclusion}
In conclusion, it has been argued that the superconductivity type is characterized by the SE at the normal-SC interface, which determines whether the system is type-I (type-II) SC corresponding to positive (negative) SE.
Superconductivity type is an integral energetic property, because response of a SC body placed into an external magnetic flux directly depends on sign of the associated energy.
Numerical results suggest that the SE is a monotonic function of the GL parameter, and therefore, there is only a single value for $\kappa$ that renders the system to the unique special case where the SC response is in a sense exactly between the two conventional types I and II.
Analytical prediction for the interaction parameters sets which render the s.e. to zero has been found by noting that in case when SE is equal to zero the thermodynamic MF and the upper critical MF coincide: $H_c=H_{c2}$.
The condition $H_c=H_{c2}$ arises on the grounds of the physical interpretation of the basic functional, and therefore, it is valid independently of whether the SC is a single component quantum fluid or a SC-SF mixture.
The analytical result for the critical GL parameter, Eq. (\ref{kappaCdens}), has been confirmed by numerical solution to the boundary value problem providing reliable predictions, as displays Table I.
It is emphasized that coexistence of the SF and the SC in a uniform mixture is possible only in a limited domain of the parameter space as show Eqs. (\ref{stability1})-(\ref{stability3}) and Fig. 1 for a special case.

The findings of the present work regarding the effect of the bare density coupling on type of superconductivity, reached here in the framework of SE of a planar interface, are equivalent to the ones reached earlier in the literature in the framework of forces between the quantized flux tubes \cite{BuckleyEtalPRC2004,AlfordGood2008,HaberSchmitt2017}.
This is not surprising because the basic energy functionals here and in \cite{AlfordGood2008} are equivalent, with a reservation that the functional in \cite{AlfordGood2008} is a special case of Eq. (\ref{FGL}).
For sufficiently large bodies, which are in the focus here, this energy can be interpreted either as (i) the energy of the surface per unit area, (ii) or as the energy of a system of superconducting interacting flux tubes, which underlie the screening SC current.
Note that the latter interpretation is relevant to the planar interface by virtue of the fact that a planar interface can be thought of as boundary of the core of a very large flux tube, due to the topological reasons and because its curvature is negligible \cite{AlfordGood2008}.
In the first picture, one finds the sign of energy of a small planar piece of the surface (per unit area), and the integral energy is simply a multiple of the result.
In the second picture, employed in \cite{AlfordGood2008}, conclusions are inferred on the basis of tube-tube interactions at zero and infinite separation.

The present numerical results are limited by a single value of the SF density ratio (the parameter $\eta=0.05$), following the earlier literature on the subject \cite{AlfordGood2008}.
However, the parameter $\eta$ depends on the total baryon density and is expected to take on values between 0 and $+\infty$.
The results reported here are relevant for a small layer with equal critical temperatures, which possibly exists in the NS core.
Within this layer, the position of the normal-SC surface is determined by thermodynamic balance between the MF at one side of the surface and the proton order parameter with the Meissner currents at the other side.
Description of superconductivity outside this layer is an important future task, which shall be crucial for conclusions on the global structure of the NS.

Understanding the links between thermodynamic coupling constants introduced here with the coupling constants used in theories valid at higher excitation energies is necessary in future studies.
One of the most important steps in this direction would be to solve the nuclear pairing problem with three-body forces and to retain the higher orders of the order parameter that would couple the pairing equations for neutrons and protons allowing for a self-consistent calculation of the both SF critical temperatures.
These tasks go beyond the scope of this work; their solution will help establish a reliable theoretical picture of superconductivity properties of nuclear matter in the core of NS.

\section{Acknowledgments}
I thank Mark Alford, Andreas Schmitt, Lev Smirnov, Matt Caplan for their valuable remarks and useful discussions.

\section{Appendix}
\subsection{Single-component nonlinear Schr$\ddot{\mathrm{o}}$dinger equation}
It is instructive to derive a compact form of the nonlinear Schr$\ddot{\mathrm{o}}$dinger equation for a single SF with a perspective to two-component generalization.
The nonlinear Sch$\ddot{\mathrm{o}}$dinger equation in a single-component case reads
\begin{equation}
\label{NLSE1}  \mathrm{i}\hbar\frac{\partial\psi}{\partial t}=-\frac{\hbar^2}{2m}\nabla^2\psi+\frac{\partial E_{\mathrm{st}}^{\mathrm{nuc}}}{\partial n}\psi.
\end{equation}
Substitution of Eq. (\ref{psi}) into Eq. (\ref{NLSE1}) yields an equivalent set of two equations,
\begin{eqnarray}
  \label{cont1} \frac{\partial n}{\partial t}+\nabla\left(n\frac{\hbar\nabla\phi}{m}\right)=0, \\
  \label{Eul1} \hbar\frac{\partial \phi}{\partial t}+\mu^{\mathrm{tot}}=0,
\end{eqnarray}
where
\begin{equation}
\label{mutot1}  \mu^{\mathrm{tot}}\equiv\frac{\partial H^{\mathrm{Schr}}}{\partial n}=\frac{\hbar^2}{2m}(\nabla\phi)^2-\frac{\hbar^2}{2m}\frac{\nabla^2\sqrt{n}}{\sqrt{n}}+\frac{\partial E_{\mathrm{st}}^{\mathrm{nuc}}}{\partial n}
\end{equation}
is a scalar, which is equal to a sum of the kinetic energy density, the quantum pressure, and the usual chemical potential in the absence of matter flows and the SF density gradient.
The Hamiltonian $H^{\mathrm{Schr}}$ is
\begin{equation}\label{Hschr}
H^{\mathrm{Schr}}=\frac{\hbar^2}{2m}|\nabla\psi|^2+E_{\mathrm{st}}^{\mathrm{nuc}}[n].
\end{equation}
The Hamiltonian  of the ideal fluid is:
\begin{eqnarray}
\nonumber && \!\!\!\!\!\!\!\!\!\!\!\!\!\!\!H^{\mathrm{id}}= n\frac{(\hbar\nabla\phi)^2}{2m}+E_{\mathrm{st}}^{\mathrm{nuc}}[n]\\
\label{Hid} &&=\frac{\hbar^2}{2m}\left(|\nabla\psi|^2-(\nabla|\psi|)^2\right)+E_{\mathrm{st}}^{\mathrm{nuc}}[n]
\end{eqnarray}
Note that the quantity $\frac{\partial E_{\mathrm{st}}^{\mathrm{nuc}}}{\partial n}$ can be identified with the chemical potential only when the SF density $n$ is equal to the total particle density.
The interpretation of Eqs. (\ref{cont1}) and (\ref{Eul1}) as the continuity equation and the Euler equation is consistent with the interpretation of $|\psi|^2$ as the probability density and $\hbar\nabla\phi$ as the specific momentum.
I emphasize that the definition of the number current in mechanics
\begin{equation}\label{J1comp}
\mathbf{J}=\frac{\hbar}{m}\frac{\psi^*\nabla\psi-c.c.}{2\mathrm{i}}\equiv n\frac{\hbar\nabla\phi}{m}
\end{equation}
can be viewed as a direct consequence of the continuity equation Eq. (\ref{cont1}).
The static (in the absence of matter flows) internal energy density of ideal fluid is a functional of the number density $n$: $E_{\mathrm{st}}^{\mathrm{nuc}}=E_{\mathrm{st}}^{\mathrm{nuc}}[n]$.
As shows Eq. (\ref{Hid}), in the single-component ideal fluid the quantity $E_{\mathrm{st}}^{\mathrm{nuc}}$ is the internal fluid energy.
The remaining part of the total energy, $n{(\hbar\nabla\phi)^2}/{2m}$ can be interpreted as the kinetic energy of the fluid flow.
The second line in Eq. (\ref{Hid}) shows that the quantum pressure is subtracted from the total quantum kinetic energy, thus, the ideal fluid model is equivalent to the nonlinear Schr$\ddot{\mathrm{o}}$dinger model without the quantum pressure.
It is convenient to rewrite the nonlinear Sch$\ddot{\mathrm{o}}$dinger equation using Eqs. (\ref{cont1}) and (\ref{Eul1}) in the following form:
\begin{equation}\label{NLSE1form}
  \mathrm{i}\hbar\frac{\partial\psi}{\partial t}=\left[\mu^{\mathrm{tot}}-\mathrm{i}\frac{\hbar}{2mn}\nabla\left(n\nabla\phi\right)\right]\psi.
\end{equation}
\subsection{Two-component Schr$\ddot{{\mathrm{o}}}$dinger model}
It is also instructive to observe a two-component generalization of the nonstationary Schr$\ddot{\mathrm{o}}$dinger equation:
\begin{equation}\label{NLSE2form}
  \mathrm{i}\hbar\frac{\partial\psi_\alpha}{\partial t}=\left[\mu^{\mathrm{tot}}_\alpha-\mathrm{i}\frac{1}{2n_\alpha}\nabla\left(n_\alpha\mathbf{v}_\alpha\right)\right]\psi_\alpha,
\end{equation}
with the flow velocities $\mathbf{v}_\alpha$ given by Eqs. (\ref{vel1}) and  (\ref{vel2}), and where
\begin{widetext}
\begin{eqnarray}
\label{muptot}&&   \mu_{p}^{\rm tot}=\frac{P_p^2}{2m}-\frac{\hbar^2}{2m}\frac{\nabla^2\sqrt{n_p}}{\sqrt{n_p}}-\frac{1}{2m}\frac{\partial n_{np}}{\partial n_{p}}\left(\mathbf{P}_p-\mathbf{P}_n\right)^2
  -\frac{\lambda_{np}}{2\sqrt{n_p}}\nabla^2\sqrt{n_n}-\frac{\mu_{np}}{4}\nabla^2 n_n+\frac{\partial E_{\mathrm{st}}^{\mathrm{nuc}}}{\partial n_p}+e\Phi,\\
 \label{muntot}&&  \mu_{n}^{\rm tot}=\frac{P_n^2}{2m}-\frac{\hbar^2}{2m}\frac{\nabla^2\sqrt{n_n}}{\sqrt{n_n}}-\frac{1}{2m}\frac{\partial n_{np}}{\partial n_{n}}\left(\mathbf{P}_p-\mathbf{P}_n\right)^2
    -\frac{\lambda_{np}}{2\sqrt{n_n}}\nabla^2\sqrt{n_p}-\frac{\mu_{np}}{4}\nabla^2 n_p+\frac{\partial E_{\mathrm{st}}^{\mathrm{nuc}}}{\partial n_n},
\end{eqnarray}
\end{widetext}
and $\Phi$ is the scalar electric potential.
The number currents in this model are given by the expressions equivalent to Eq. (\ref{currdef}) with Eqs. (\ref{vel1}) and (\ref{vel2}):
\begin{widetext}
\begin{eqnarray}
\label{Jp}&& \mathbf{J}_p=\frac{\hbar}{m}\frac{\psi_p^*\nabla\psi_p-c.c.}{2\mathrm{i}} -\frac{\hbar}{2\mathrm{i}m}n_{np}\left(\frac{\psi_p^*\nabla\psi_p-c.c.}{|\psi_p|^2}-\frac{\psi_n^*\nabla\psi_n-c.c.}{|\psi_n|^2}\right), \\
\label{Jn} && \mathbf{J}_n=\frac{\hbar}{m}\frac{\psi_n^*\nabla\psi_n-c.c.}{2\mathrm{i}}
 -\frac{\hbar}{2\mathrm{i}m}n_{np}\left(\frac{\psi_n^*\nabla\psi_n-c.c.}{|\psi_n|^2}-\frac{\psi_p^*\nabla\psi_p-c.c.}{|\psi_p|^2}\right). \end{eqnarray}
\end{widetext}
In the linear regime and in the absence of the MF, this model is equivalent to the formalism developed in \cite{KobyakovPethick2017} and \cite{KobyakovEtal2017}.

\end{document}